UDC 541.2

# Superheavy elements: existence, classification and experiment

**V.A.Kostyghin[1], V.M.Vaschenko[2], Ye.A.Loza[2]**

[1] *"Akvatekhinzhiniring", Cherkassy, Ukraine, e-mail:kostygin558@mail.ru*

[2] *State Ecological academy for post-graduate education and management, Kyiv, Ukraine, e-mail:daniilko@mail.ru*

This paper proposes spatial periodic table developed based on classic electron shell structure model. The periodic table determines location and chemical properties of superheavy elements. 14 new long-living superheavy elements found by Proton-21 laboratory and one long-living superheavy element found by A.Marinov were identified.

## Introduction

At present there are 118 elements known [1]. However, there are only 104 of them well-studied. The superheavy nuclei existence problem and also nuclones collective states problems (super-nuclei, nuclear associations, multi-nuclei clusters, nuclear condensate etc.) long ago have become subject of experimental and theoretical research [2-7]. This problem is specially important for modern astrophysical concepts [8-10].

The main branch of experimental research of superheavy nuclei isotopes and exotic isotopes of light nuclei is a new elements synthesis by different accelerator with latter investigation of nuclei collision results. However this method has two important and unsolved problems [11, 12].

The first one is the excess energy of the synthesized nucleus. The energy required to overcome the Coulomb barrier during the nuclei collision transforms into internal energy of the newly formed nucleus and it is usually enough for instant nuclei fission, because the internal clusters of the nuclei have energy over Coulomb barrier. This leads to a very complex experimental task of "discharge" of excess energy by high-energy particle radiation - gamma-quantum, neutron, positron, proton, alpha-particle [13] etc. As a result of excess energy the synthesized nuclei lifetime may be $10^{-22}$s and less [14, 15].

The second problem is a neutron deficiency of the synthesized elements [13, 16]. The isotopes stability curve manifests non-linear increase of required neutrons for nucleus stabilization by proton quantity increase. Therefore it is almost impossible to obtain a stable superheavy nucleus isotope (so-called islands of stability) from usual combination nuclei because all the heavy nuclei synthesized by collision of lighter nuclei will have neutron deficiency. The experiments on nuclei collision are usually made on neutron excess nuclei, such as several lead isotopes. Thou this doesn't guarantee the synthesized nucleus stability, because its stable isotope requires a specific neutron-proton ratio (usually having a few possible variants) and while pursing heavier nuclei synthesis this neutron excess becomes insufficient.

Both mentioned problems of superheavy nuclei synthesis in modern accelerators lead to extremely small lifetimes of the synthesized nuclei and underestimate of their possible lifetime in normal conditions.

Additional experimental ambiguity is uniqueness of successful superheavy nuclei syntheses events. I.e. there were only 3 cases of 118th element registered by 2011 [17]. Not only this restricts speaking of its physical-and-chemical properties but also brings concerns on purposefulness and experimental reliability of such experiment.

Moreover very small lifetimes (less than a second) of the obtained nuclei don't enable studying electron shells configuration, because they have time to form and relaxate and therefore there is not enough time for obtaining optical spectrum and chemical properties.

Theoretical investigations of superheavy nuclei is made by quantum modeling. However, strong force behavior at small distances is unknown. Moreover, quantum many-body problem has no rigor solution at present. Therefore such modeling often yields contradictory results. Moreover diversity and nonconformity of nuclei models leads to significant errors even for well-studied light nuclei.

In this paper, understanding all the problems of superheavy nuclei and electron shells behavior theory we try a simplified approach based on Madelung (Klechkovsky) rule which yield stable accuracy for known chemical elements and predictable deviation scale [18].

## Superheavy elements mass determination problem

For experimental search for superheavy elements except their charge, electron shell structure and chemical properties the data on their atomic mass is also required, which consists of neutron and proton total mass "minus" interaction energy. Then in case of mass-spectrometric or tracks investigations some events may be positively identified as relating to superheavy elements.

Another problem is that the atomic mass dependance on atomic number is ambiguous and one chemical element may have multiple stable isotopes, determined by a complex and partially unknown principle.

One of the most simple approach we use in this paper is «freezing» of neutron to atomic mass ratio of relatively stable isotopes having half-life time over 100 days. For stability island isotopes this ratio is 0.609±0.003 and asymptotically approaches "the golden section" 0.618.

Therefore, considering statistical deviations we may suggest that atomic mass is 2.536 to 2.574 times greater than nucleus charge. In this paper we use value 2.54, also used by other authors [1, 4]. If suggestion of static neutron-to-proton ratio is true then statistical error estimate of nucleus mass at atomic number 550 will be around 12 a.m.u., which is quiet well for modern mass-spectrometric investigations.

Such dependency is incorrect for elements with atomic number over 1000 because it yields statistical error over 20 a.m.u.

However, we must note that such dependency in general case is incorrect. As said above, the nuclei mass dependency on its charge is ambiguous and non-linear function and just for the next (linear) approximation the estimate leads to almost doubling the estimated nucleus mass for element number 550 and it is almost three times greater for 1000th element comparing to static model.

Therefore it is quiet unreasonable to estimate superheavy nuclei mass with atomic number over 250 which has difference between static and linear model at +34.7%. More exact theoretical-and-experimental nucleus stability curve is required.

Let's stress that difference of just 1 proton may lead to significant chemical and optical properties change. As a consequence, the most calculations in this paper made in static model should be considered as example of the model application.

## Maximal atomic number

Another problem considering superheavy nuclei investigation is the maximum possible quasi-stable elements in the nature. This question mainly relates to nuclones collective states and association problem, strong force peculiarities and many-body problem as said above. Here we just give the results of different authors.

Let's note, we investigate here only atoms in their classic understanding (i.e. a relatively small quasi-spherical homogeneous nucleus with multi-electron shells). We do not consider so-called «bubble» nuclei, superdense nuclei, supercharged nuclei, neutron nuclei, hadron nuclei [2] etc.

As soon as all these estimates are based on approximate nuclei structure models none of them may be considered ultimate.

Kapustinsky and Ozhigov earlier predicted the total maximum number of elements in the periodic table to be 120 [19, 20].

The estimate made according to relativistic Dirac equation yields maximal amount of 137 elements [21], while 138th element is predicted to have imaginary energy. The same maximal quantity of elements is obtained by classic Bohr model, when electron speed at lowest orbit in hydrogen-like ion becomes greater than speed of light.

Some nucleus models predict equality of electromagnetic and strong forces at around Z=150. Khazan based on method of rectangular hyperbolas suggested absolute maximum of possible nuclei charge 155 with atomic mass 411.663243 [22, 23].

The periodic table symmetry considerations yields 170 as maximum of elements possible in nature [24, 25]. Another symmetry considerations provide for 170 or 220 elements [18] (however, the total number of elements is not strictly limited to this number, 292, 390, 518, 680 and more satisfying symmetry conditions are possible).

Other authors predict total number of elements to be either 172 [26] or 173 [27] when according to relativistic Dirac equation atom ground state energy becomes greater than double electron rest energy (i.e. spontaneous beta-decay of the nuclei is simultaneously triggered in case ground state is free).

V.A. Kostyghin based on electron orbit and nuclei radii comparison and electron orbital properties suggests absolute maximal numbers of elements to be 558 [28].

In this approach the maximal number of elements is derived from the following theoretical considerations.

The radius of first orbital for hydrogen-like ion is determined by:

$$R_{ion} = \frac{h^2 \varepsilon_0}{\pi m_e e^2 Z} \quad (1)$$

where h = 6.626·$10^{-34}$ $m^2$ kg/s is Planck constant, $\varepsilon_0$ = 8.854·$10^{-12}$ F·$m^{-1}$ - dielectric vacuum permittivity, $m_e$ = 9.109·$10^{-31}$ kg is electron mass, e = 1.602·$10^{-19}$ C is the electron charge, π = 3.14..., and Z is the nucleus charge.

On the other hand, by rough approximation we may present atom's nucleus as a sphere, then its mass is defined by:

$$M = (4/3)\ \pi \cdot \rho \cdot R_{nucleus}^3 \quad (2)$$

where ρ and $R_{nucleus}$ is nucleus density and radius, correspondingly. In this paper we consider $\rho = m_p \cdot/(4\pi/3)/r_0^3$, where $m_p$ = 1.673·$10^{-27}$ kg is proton mass and $r_0$ = 1.25·$10^{-15}$ m is nucleus radius constant [29] (where deviations up to 20% from this value are observed for known nuclei). Therefore, the mean nucleus density value is ρ = 2.04·$10^{17}$ kg/$m^3$ (while maximum nucleus density for known nuclei is 3.99·$10^{17}$ kg/$m^3$ and minimum is 1.18·$10^{17}$ kg/$m^3$).

On the other hand we may approximately calculate M by total amount of protons and neutrons in the nucleus:

$$M = Z \cdot j \cdot m_p \quad (3)$$

where $m_p$ is proton mass and j = 2.54 is a linear empirical coefficient discussed earlier.

In case $R_{ion} = R_{nucleus}$ the elements will undertake an inevitable rapid stability drop, therefore, considering (2) and (3) it defines the final element to be considered in our model:

$$Z = \left( \frac{4}{3} \frac{\pi\rho}{jm_p} \left( \frac{h^2 \varepsilon_0}{\pi m_e e^2} \right)^3 \right)^{1/4} \qquad (4)$$

Finally we obtain solution $Z_{max} = 2338$ for mean nuclei density. Considering final closed shell element in the periodic table we assume $Z_{max} = 2310$.

Considering possible nucleus density deviation for known nuclei, the maximum possible value is 2756 (i.e. the final closed-shell element is 2598th and 23 periods in the periodic table) and minimum value is 2039 (the final closed-shell element is 3022th and 21 periods in the periodic table) according to nuclear density deviations.

Therefore we consider 2310 elements as base for the spatial periodic table.

## Spatial periodic table

Today there are many different variants of the periodic table extrapolation to include superheavy nuclei. Earlier on group theory basis, group SO(4, 2) was used to describe the periodic table [30, 31]. In this paper we propose advanced model of SO(4, 2) group in graph table form. The main goal of this model is to show the limits and the structure of the periodic table, to forecast properties of unknown elements and to propose a method for search for superheavy nuclei.

The full theoretical basis for spatial periodic table construction was published earlier [28, 32, 33].

The result of its construction is given below.

In this way we visualize every chemical element with a 3D cube corresponding to its electron structure build a spatial periodic table. Practically this model is a volume graph of atomic principal quantum numbers.

Let us consider the model in detail.

The 1st period is composed of Hydrogen and Helium, their electron structure is $1s^1$ and $1s^2$ (electron structure of all elements is taken from [34]).

The 2nd period consists of Li, Be, - s-elements and B, C, N, O, F, Ne, - p-elements.

The 3rd period: Na, Mg, Al, Si, P, S, Cl, Ar. s-elements are placed under s-elements, and p-elements are placed under p-elements.

While building 4th period when we come to d-elements there is a turn and when we come to p-elements of 4th period Ga, Ge, As, Se, Br, Kr - there is another turn. These elements are placed parallel to p-elements of 2nd and 3rd periods.

Similar to 4th period we fill the 5th period.

There are lanthanides in the sixth period that "follow" the same direction as p-elements. But, starting from lanthanides there are some new peculiarities. First, lanthanides form a loop. The first loop side consists of: La, Ce, Pr, Nd, Pm, Sm, Eu, Gd. The second loop side consists of: Tb, Dy, Ho, Er, Tm, Yb, Lu. Second, the cells are formed with two elements: La/Ce and Hf/Lu.

The 7th (unfinished) period contains actinides that are placed under lathanides forming a similar loop, e.g. Ac matches Th, The loop turn takes place at Cm and Bk, and Lr shares a single cell with Ku.

The 8th and 9th periods are characterized by a new element class: the g-elements. They form loop parallel to d-elements, there are 4 matched elements pairs with f-elements, i.e. there are 8 matched elements.

The 10th and 11th periods each contains 22 h-elements, they form a complex loop parallel to f-elements, "crossing" with g-elements. There are 6 pairs of "double" elements.

12th and 13th periods each contains 26 i-elements filling all the remaining internal cells between f, d, p-elements.

14th and 15th periods each contains 30 j-elements (let's note that some notation systems do not use letter j for electron shells in order not to be confused with i). 16th and 17th periods each contains 34 k-elements. 18th and 19th periods each contains 38 v-elements. 20th and 21st both include 42 w-elements. And the 22nd period has 46 t-elements.

Therefore, the spatial model contains 2310 elements and consists of 22 periods.

## Method of search for and identification of superheavy elements

It is known, that with increase of elements charge their chemical properties differ less, because the chemical properties are provided by external orbitals but the internal orbitals are being filled. I.e. the admixtures of superheavy elements may almost inseparably exist in light elements and could be identified by standard mass-spectrometric means. However, there is a method uncertainty of expected superheavy nucleus peak identification. This is either superheavy nucleus either molecular ion of a specimen impurity.

The obtained spatial model of periodic table allows obtaining mass numbers of superheavy elements, that can be identified is the mass spectrum of a specific element as nuclei and not molecular ions.

Fig.1 presents the outer view of spatial periodic table up to Z=558. And fig.2 presents outer view of spatial periodic model up to Z=2310.

Fig. 3-24 present internal "sections" i.e. the periods of the model for all 22 periods.

Let us give the procedure of identification of superheavy elements nuclei using the spatial periodic table (up to 558th element). E.g. we take Cu subgroup. this subgroup includes the following elements: Cu, Ag, Au, 111, 161, 132, 133, 211, 182, 183. Multiplying 2.54 by nucleus charge we obtain mass numbers of Cu subgroup: (starting from 111th) 218.94; 408.94; 335.28; 337.82; 535.94; 462.28; 464.82.

Then we have to obtain Cu specimen mass spectrum and analyze it. If it will have peaks corresponding to the found atomic numbers, then there is a high probability that this is a super-heavy nucleus peak, not a molecular ion, because the probability of a specific number combination is low when explaining given experimental peaks as molecular. Similar procedure may be also done for other known elements. Some numbers will cross over, therefore increasing probability of superheavy nuclei detection. The full information on availability of superheavy elements (including 13 periods) is given in the Table 1.

However, we should note that this method will only work for superheavy nuclei rich sample with superheavy nuclei concentration over $10^{-5}$. Thou according to Marinov estimates in normal chemical compounds their concentration is about $10^{-9}$ to $10^{-12}$ [35, 36] which is far less than obtainable sample cleanness which may lead to misinterpretation of molecular ions with different isotopes as superheavy nuclei. So, much better way is to find possible physical and/or chemical reactions enabling "enrichment" of superheavy nuclei content in the sample. Another way is obtaining superheavy nuclei-rich samples (e.g. synthesizing stable superheavy nuclei by some "cold" nuclear reaction).

However, we should point out that just as regular ores, some rocks may be rich in superheavy nuclei, thou their concentration is expected to be far less than that of lighter elements. E.g. Migdal proposed to search for astrophysically-generagted superheavy elements in lunar rock, meteorites and cosmic rays [2, 10].

Synthesis in nuclear explosions [8, 13] eliminates both synthesis problems providing for automatic energy discharge and automatic neutron sufficiency, but generating sample collection problem.

**Table 1. Superheavy elements groups**

| | **La Ce Th** | **Hf Lu** | **V Nb Ta** | **Cr Mo W** | **Mn Te Re** | **Fe Ru Os** | **Co Rh Ir** | **Ni Pd Pt** | **Cu Ag Au** |
|---|---|---|---|---|---|---|---|---|---|
| nuclei charge | 121 | 154 | 105 | 106 | 107 | 108 | 109 | 110 | 111 |
| atomic number | 307.34 | 391.16 | 266.7 | 269.24 | 271.78 | 274.32 | 276.86 | 279.4 | 281.94 |
| nuclei charge | 122 | 153 | 155 | 156 | 177 | 158 | 159 | 160 | 161 |
| atomic number | 309.88 | 388.62 | 393.7 | 396.24 | 449.58 | 401.32 | 403.86 | 406.4 | 408.94 |
| nuclei charge | 123 | 152 | 126 | 127 | 128 | 129 | 130 | 131 | 132 |
| atomic number | 312.42 | 386.08 | 320.04 | 322.58 | 325.12 | 327.66 | 330.2 | 332.74 | 335.28 |
| nuclei charge | 124 | 125 | 139 | 138 | 137 | 136 | 135 | 134 | 133 |
| atomic number | 314.96 | 317.5 | 353.06 | 350.52 | 347.98 | 345.44 | 342.9 | 340.36 | 337.82 |
| nuclei charge | 142 | 151 | 205 | 206 | 207 | 208 | 209 | 210 | 211 |
| atomic number | 360.68 | 383.54 | 520.7 | 523.24 | 525.78 | 528.32 | 530.86 | 533.4 | 535.94 |
| nuclei charge | 141 | 140 | 176 | 177 | 178 | 179 | 180 | 181 | 182 |
| atomic number | 358.14 | 355.60 | 447.04 | 449.58 | 452.12 | 454.66 | 457.2 | 459.74 | 462.28 |
| nuclei charge | 143 | 150 | 189 | 188 | 187 | 186 | 185 | 184 | 183 |
| atomic number | 363.22 | 381.00 | 480.06 | 477.52 | 474.98 | 472.44 | 469.9 | 467.36 | 464.82 |
| nuclei charge | 144 | 149 | 277 | 278 | 279 | 280 | 281 | 282 | 283 |
| atomic number | 365.76 | 378.46 | 703.58 | 706.12 | 708.66 | 711.2 | 713.74 | 716.28 | 718.82 |
| nuclei charge | 145 | 148 | 248 | 249 | 250 | 251 | 252 | 253 | 254 |
| atomic number | 368.30 | 375.92 | 629.92 | 632.46 | 635 | 637.54 | 640.08 | 642.62 | 645.16 |
| nuclei charge | 146 | 147 | 226 | 237 | 259 | 258 | 257 | 256 | 255 |
| atomic number | 370.84 | 373.38 | 574.04 | 601.98 | 657.86 | 655.32 | 652.78 | 650.24 | 647.7 |
| nuclei charge | 171 | 203 | 261 | 260 | 240 | 247 | 353 | 354 | 355 |
| atomic number | 434.34 | 515.62 | 662.94 | 660.4 | 609.6 | 627.38 | 896.62 | 899.16 | 901.7 |
| nuclei charge | 172 | 204 | 227 | 236 | 241 | 246 | 324 | 325 | 326 |
| atomic number | 436.88 | 518.16 | 576.58 | 599.44 | 612.14 | 624.84 | 822.96 | 825.5 | 828.04 |
| nuclei charge | 173 | 202 | 228 | 235 | 242 | 245 | 329 | 328 | 327 |
| atomic number | 439.42 | 513.08 | 579.12 | 596.9 | 614.68 | 622.3 | 835.66 | 833.12 | 830.58 |
| nuclei charge | 174 | 175 | 229 | 234 | 243 | 244 | 451 | 452 | 453 |
| atomic number | 441.96 | 444.50 | 581.66 | 594.36 | 617.22 | 619.76 | 1145.54 | 1148.08 | 1150.62 |
| nuclei charge | 192 | 201 | 230 | 233 | 351 | 352 | 422 | 423 | 424 |
| atomic number | 487.68 | 510.54 | 584.2 | 591.82 | 891.54 | 894.08 | 1071.88 | 1074.42 | 1076.96 |
| nuclei charge | 191 | 190 | 231 | 232 | 322 | 323 | 427 | 426 | 425 |
| atomic number | 485.14 | 482.60 | 586.74 | 589.28 | 817.88 | 820.42 | 1084.58 | 1082.04 | 1079.5 |
| nuclei charge | 193 | 200 | 349 | 350 | 310 | 330 | 377 | 378 | 379 |
| atomic number | 490.22 | 508.00 | 886.46 | 889 | 787.4 | 838.2 | 957.58 | 960.12 | 962.66 |
| nuclei charge | 194 | 199 | 320 | 321 | 331 | 319 | 382 | 361 | 380 |
| atomic number | 492.76 | 505.46 | 812.8 | 815.34 | 840.74 | 810.26 | 970.28 | 916.94 | 965.2 |
| nuclei charge | 195 | 198 | 298 | 309 | 311 | 318 | 391 | 392 | 393 |
| atomic number | 495.30 | 502.92 | 756.92 | 784.86 | 789.94 | 807.72 | 993.14 | 995.68 | 998.22 |
| nuclei charge | 196 | 197 | 333 | 332 | 312 | 317 | 396 | 395 | 394 |
| atomic number | 497.84 | 500.38 | 845.82 | 843.28 | 792.48 | 805.18 | 1005.84 | 1003.3 | 1000.76 |
| nuclei charge | 221 | 276 | 299 | 308 | 313 | 316 | 549 | 550 | 551 |
| atomic number | 561.34 | 701.04 | 759.46 | 782.32 | 795.02 | 802.64 | 1394.46 | 1397 | 1399.54 |
| nuclei charge | 222 | 275 | 300 | 307 | 314 | 450 | 520 | 521 | 522 |
| atomic number | 563.88 | 698.50 | 762 | 779.78 | 797.56 | 1143 | 1320.8 | 1323.34 | 1325.88 |
| nuclei charge | 223 | 274 | 301 | 306 | 315 | 421 | 525 | 524 | 523 |
| atomic number | 566.42 | 695.96 | 764.54 | 777.24 | 800.1 | 1069.34 | 1333.5 | 1330.96 | 1328.42 |
| nuclei charge | 224 | 225 | 302 | 305 | 449 | 428 | 475 | 476 | 477 |
| atomic number | 568.96 | 571.50 | 767.08 | 774.7 | 1140.46 | 1087.12 | 1206.5 | 1209.04 | 1211.58 |
| nuclei charge | 264 | 273 | 303 | 304 | 408 | 417 | 480 | 479 | 478 |
| atomic number | 670.56 | 693.42 | 769.62 | 772.16 | 1036.32 | 1059.18 | 1219.2 | 1216.66 | 1214.12 |
| nuclei charge | 263 | 262 | 447 | 448 | 420 | 376 | 489 | 490 | 491 |
| atomic number | 668.02 | 665.48 | 1135.38 | 1137.92 | 1066.8 | 955.04 | 1242.06 | 1244.6 | 1247.14 |
| nuclei charge | 265 | 272 | 370 | 407 | 409 | 416 | 494 | 493 | 492 |
| atomic number | 673.10 | 690.88 | 939.8 | 1033.78 | 1038.86 | 1056.64 | 1254.76 | 1252.22 | 1249.68 |
| nuclei charge | 266 | 271 | 418 | 419 | 429 | 383 | - | - | - |
| atomic number | 675.64 | 688.34 | 1061.72 | 1064.26 | 1089.66 | 972.82 | - | - | - |
| nuclei charge | 267 | 270 | 371 | 406 | 410 | 415 | - | - | - |
| atomic number | 678.18 | 685.80 | 942.34 | 1031.24 | 1041.4 | 1054.1 | - | - | - |
| nuclei charge | 268 | 269 | 431 | 430 | 375 | 390 | - | - | - |
| atomic number | 680.72 | 683.26 | 1094.74 | 1092.2 | 952.5 | 990.6 | - | - | - |
| nuclei charge | 293 | 348 | 372 | 405 | 411 | 414 | - | - | - |
| atomic number | 744.22 | 883.92 | 944.88 | 1028.7 | 1043.94 | 1051.56 | - | - | - |
| nuclei charge | 294 | 347 | 373 | 374 | 384 | 397 | - | - | - |
| atomic number | 746.76 | 881.38 | 947.42 | 949.96 | 975.36 | 1008.38 | - | - | - |
| nuclei charge | 295 | 346 | 399 | 404 | 412 | 548 | - | - | - |
| atomic number | 749.30 | 878.84 | 1013.46 | 1026.16 | 1046.48 | 1391.92 | - | - | - |
| nuclei charge | 296 | 297 | 386 | 385 | 389 | 519 | - | - | - |
| atomic number | 751.84 | 754.38 | 980.44 | 977.9 | 988.06 | 1318.26 | - | - | - |
| nuclei charge | 336 | 345 | 400 | 403 | 413 | 526 | - | - | - |
| atomic number | 853.44 | 876.30 | 1016 | 1023.62 | 1049.02 | 1336.04 | - | - | - |
| nuclei charge | 335 | 334 | 387 | 388 | 398 | 515 | - | - | - |
| atomic number | 850.90 | 848.36 | 982.98 | 985.52 | 1010.92 | 1308.1 | - | - | - |
| nuclei charge | 337 | 344 | 401 | 402 | 547 | 474 | - | - | - |
| atomic number | 855.98 | 873.76 | 1018.54 | 1021.08 | 1389.38 | 1203.96 | - | - | - |
| nuclei charge | 338 | 343 | 545 | 546 | 506 | 514 | - | - | - |
| atomic number | 858.52 | 871.22 | 1384.3 | 1386.84 | 1285.24 | 1305.56 | - | - | - |
| nuclei charge | 339 | 342 | 468 | 505 | 518 | 481 | - | - | - |
| atomic number | 861.06 | 868.68 | 1188.72 | 1282.7 | 1315.72 | 1221.74 | - | - | - |
| nuclei charge | 340 | 341 | 516 | 517 | 507 | 513 | - | - | - |
| atomic number | 863.60 | 866.14 | 1310.64 | 1313.18 | 1287.78 | 1303.02 | - | - | - |
| nuclei charge | 365 | 446 | 469 | 504 | 527 | 488 | - | - | - |
| atomic number | 927.10 | 1132.84 | 1191.26 | 1280.16 | 1338.58 | 1239.52 | - | - | - |
| nuclei charge | 366 | 445 | 529 | 528 | 508 | 512 | - | - | - |
| atomic number | 929.64 | 1130.30 | 1343.66 | 1341.12 | 1290.32 | 1300.48 | - | - | - |
| nuclei charge | 367 | 444 | 470 | 503 | 473 | 495 | - | - | - |
| atomic number | 932.18 | 1127.76 | 1193.8 | 1277.62 | 1201.42 | 1257.3 | - | - | - |
| nuclei charge | 368 | 369 | 471 | 472 | 509 | - | - | - | - |
| atomic number | 934.72 | 937.26 | 1196.34 | 1198.88 | 1292.86 | - | - | - | - |
| nuclei charge | 434 | 443 | 497 | 502 | 482 | - | - | - | - |
| atomic number | 1102.36 | 1125.22 | 1262.38 | 1275.08 | 1224.28 | - | - | - | - |
| nuclei charge | 433 | 432 | 484 | 483 | 510 | - | - | - | - |
| atomic number | 1099.82 | 1097.28 | 1229.36 | 1226.82 | 1295.4 | - | - | - | - |
| nuclei charge | 435 | 442 | 498 | 501 | 487 | - | - | - | - |
| atomic number | 1104.90 | 1122.68 | 1264.92 | 1272.54 | 1236.98 | - | - | - | - |
| nuclei charge | 436 | 441 | 485 | 486 | 511 | - | - | - | - |
| atomic number | 1107.44 | 1120.14 | 1231.9 | 1234.44 | 1297.94 | - | - | - | - |
| nuclei charge | 437 | 440 | 499 | 500 | 496 | - | - | - | - |
| atomic number | 1109.98 | 1117.60 | 1267.46 | 1270 | 1259.84 | - | - | - | - |
| nuclei charge | 438 | 439 | - | - | - | - | - | - | - |
| atomic number | 1112.52 | 1115.06 | - | - | - | - | - | - | - |
| nuclei charge | 463 | 544 | - | - | - | - | - | - | - |
| atomic number | 1176.02 | 1381.76 | - | - | - | - | - | - | - |
| nuclei charge | 464 | 543 | - | - | - | - | - | - | - |
| atomic number | 1178.56 | 1379.22 | - | - | - | - | - | - | - |
| nuclei charge | 465 | 542 | - | - | - | - | - | - | - |
| atomic number | 1181.10 | 1376.68 | - | - | - | - | - | - | - |
| nuclei charge | 466 | 467 | - | - | - | - | - | - | - |
| atomic number | 1183.64 | 1186.18 | - | - | - | - | - | - | - |
| nuclei charge | 532 | 541 | - | - | - | - | - | - | - |
| atomic number | 1351.28 | 1374.14 | - | - | - | - | - | - | - |
| nuclei charge | 531 | 530 | - | - | - | - | - | - | - |
| atomic number | 1348.74 | 1346.20 | - | - | - | - | - | - | - |
| nuclei charge | 533 | 540 | - | - | - | - | - | - | - |
| atomic number | 1353.82 | 1371.60 | - | - | - | - | - | - | - |
| nuclei charge | 534 | 539 | - | - | - | - | - | - | - |
| atomic number | 1356.36 | 1369.06 | - | - | - | - | - | - | - |
| nuclei charge | 535 | 538 | - | - | - | - | - | - | - |
| atomic number | 1358.90 | 1366.52 | - | - | - | - | - | - | - |
| nuclei charge | 536 | 537 | - | - | - | - | - | - | - |
| atomic number | 1361.44 | 1363.98 | - | - | - | - | - | - | - |

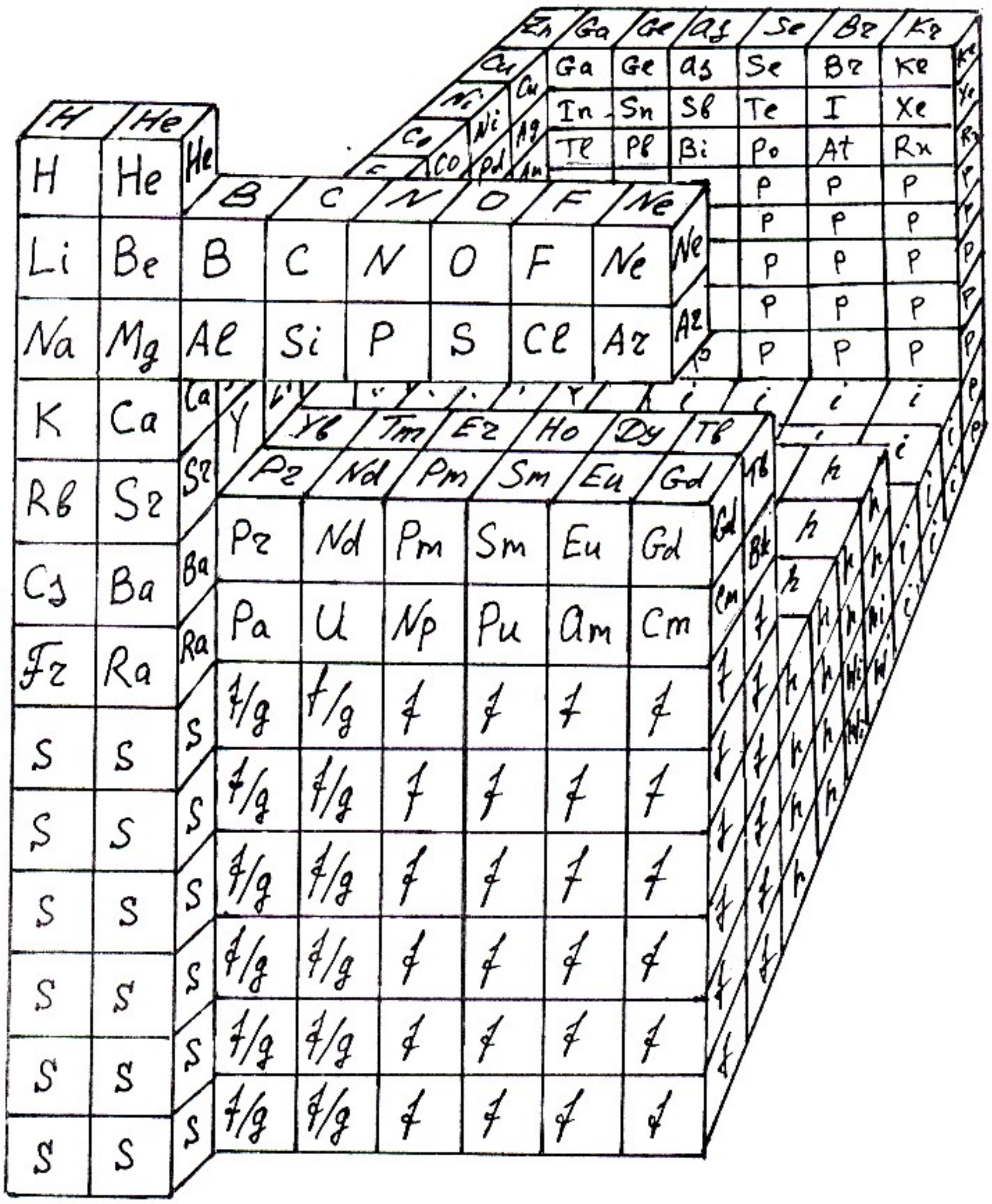

**Fig. 1. Outer view of the spatial periodic table model up to 13th period**

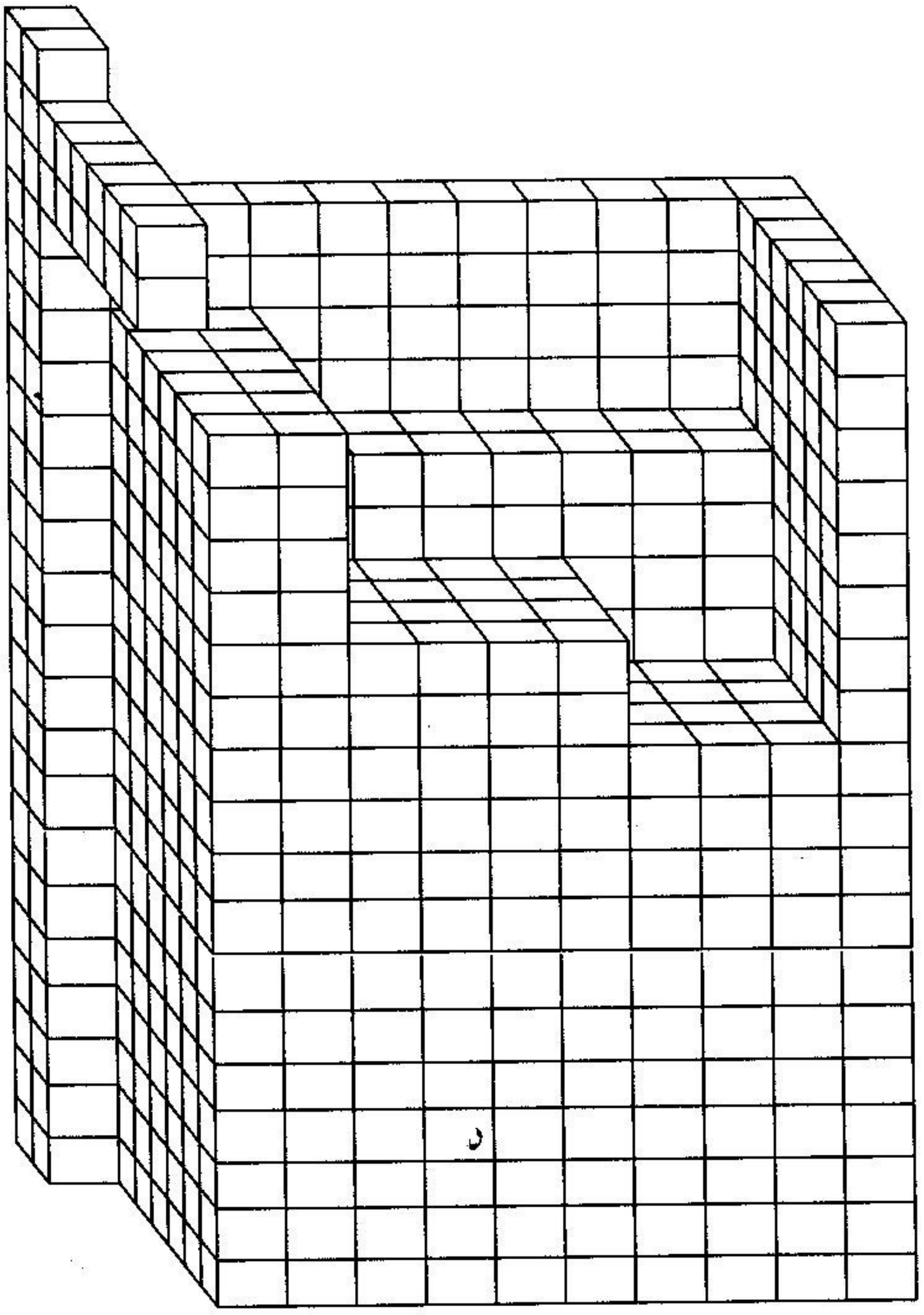

**Fig. 2. Outer view of the spatial periodic table model up to 22nd period**

| | | | | | | | | | | |
|---|---|---|---|---|---|---|---|---|---|---|
| $\mathrm{H}^1$ | | | | | | | | | | |
| $\mathrm{He}^2$ | | | | | | | | | | |
| | | | | | | | | | | |
| | | | | | | | | | | |
| | | | | | | | | | | |
| | | | | | | | | | | |
| | | | | | | | | | | |
| | | | | | | | | | | |

**Fig.3. Period 1**

| | | | | | | | | | | |
|---|---|---|---|---|---|---|---|---|---|---|
| $Li^3$ | | | | | | | | | | |
| $Be^4$ | | | | | | | | | | |
| $B^5$ | | | | | | | | | | |
| $C^6$ | | | | | | | | | | |
| $N^7$ | | | | | | | | | | |
| $O^8$ | | | | | | | | | | |
| $F^9$ | | | | | | | | | | |
| $Ne^{10}$ | | | | | | | | | | |

**Fig.4. Period 2**

| | | | | | | | | | | |
|---|---|---|---|---|---|---|---|---|---|---|
| $Na^{11}$ | | | | | | | | | | |
| $Mg^{12}$ | | | | | | | | | | |
| $Al^{13}$ | | | | | | | | | | |
| $Si^{14}$ | | | | | | | | | | |
| $P^{15}$ | | | | | | | | | | |
| $S^{16}$ | | | | | | | | | | |
| $Cl^{17}$ | | | | | | | | | | |
| $Ar^{18}$ | | | | | | | | | | |

**Fig.5. Period 3**

| | | | | | | | | | | |
|---|---|---|---|---|---|---|---|---|---|---|
| $K^{19}$ | | | | | | | | | | |
| $Ca^{20}$ | $Sc^{21}$ | $Ti^{22}$ | $V^{23}$ | $Cr^{24}$ | $Mn^{25}$ | $Fe^{26}$ | $Co^{27}$ | $Ni^{28}$ | $Cu^{29}$ | $Zn^{30}$ |
| | | | | | | | | | | $Ga^{31}$ |
| | | | | | | | | | | $Ge^{32}$ |
| | | | | | | | | | | $As^{33}$ |
| | | | | | | | | | | $Se^{34}$ |
| | | | | | | | | | | $Br^{35}$ |
| | | | | | | | | | | $Kr^{36}$ |

**Fig.6. Period 4**

| | | | | | | | | | | |
|---|---|---|---|---|---|---|---|---|---|---|
| $Rb^{37}$ | | | | | | | | | | |
| $Sr^{38}$ | $Y^{39}$ | $Zr^{40}$ | $Nb^{41}$ | $Mo^{42}$ | $Te^{43}$ | $Ru^{44}$ | $Rh^{45}$ | $Pd^{46}$ | $Ag^{47}$ | $Cd^{48}$ |
| | | | | | | | | | | $In^{49}$ |
| | | | | | | | | | | $Sn^{50}$ |
| | | | | | | | | | | $Sb^{51}$ |
| | | | | | | | | | | $Te^{52}$ |
| | | | | | | | | | | $I^{53}$ |
| | | | | | | | | | | $Xe^{54}$ |

**Fig.7. Period 5**

| | | | | | | | | | | |
|---|---|---|---|---|---|---|---|---|---|---|
| Cs$^{55}$ | | | | | | | | | | |
| Ba$^{56}$ | La$^{57}$<br>Ce$^{58}$ | Hf$^{72}$<br>Lu$^{71}$ | Ta$^{73}$ | W$^{74}$ | Re$^{75}$ | Os$^{76}$ | Ir$^{77}$ | Pt$^{78}$ | Au$^{79}$ | Hg$^{80}$ |
| | Pr$^{59}$ | Yb$^{70}$ | | | | | | | | Tl$^{81}$ |
| | Nd$^{60}$ | Tm$^{69}$ | | | | | | | | Pb$^{82}$ |
| | Pm$^{61}$ | Er$^{68}$ | | | | | | | | Bi$^{83}$ |
| | Sm$^{62}$ | Ho$^{67}$ | | | | | | | | Po$^{84}$ |
| | Eu$^{63}$ | Dy$^{66}$ | | | | | | | | At$^{85}$ |
| | Gd$^{64}$ | Tb$^{65}$ | | | | | | | | Rn$^{86}$ |

**Fig.8. Period 6**

| | | | | | | | | | | |
|---|---|---|---|---|---|---|---|---|---|---|
| $Fr^{87}$ | | | | | | | | | | |
| $Ra^{88}$ | $Ac^{89}$ $Th^{90}$ | $Ku^{104}$ $Lr^{103}$ | $d^{105}$ | $d^{106}$ | $d^{107}$ | $d^{108}$ | $d^{109}$ | $d^{110}$ | $d^{111}$ | $d^{112}$ |
| | $Pa^{91}$ | $No^{102}$ | | | | | | | | $p^{113}$ |
| | $U^{92}$ | $Md^{101}$ | | | | | | | | $P^{114}$ |
| | $Np^{93}$ | $Fm^{100}$ | | | | | | | | $p^{115}$ |
| | $Pu^{94}$ | $Es^{99}$ | | | | | | | | $p^{116}$ |
| | $Am^{95}$ | $Cf^{98}$ | | | | | | | | $p^{117}$ |
| | $Cm^{96}$ | $Bk^{97}$ | | | | | | | | $p^{118}$ |

**Fig.9. Period 7**

| | | | | | | | | | | |
|---|---|---|---|---|---|---|---|---|---|---|
| $s^{119}$ | | | | | | | | | | |
| $s^{120}$ | $d^{121}$ $f^{122}$ | $d^{154}$ $f^{153}$ | $d^{155}$ | $d^{156}$ | $d^{177}$ | $d^{158}$ | $d^{159}$ | $d^{160}$ | $d^{161}$ | $d^{162}$ |
| | $f^{123}$ $g^{124}$ | $f^{152}$ $g^{125}$ | $g^{126}$ | $g^{127}$ | $g^{128}$ | $g^{129}$ | $g^{130}$ | $g^{131}$ | $g^{132}$ | $p^{163}$ |
| | $f^{142}$ $g^{141}$ | $f^{151}$ $g^{140}$ | $g^{139}$ | $g^{138}$ | $g^{137}$ | $g^{136}$ | $g^{135}$ | $g^{134}$ | $g^{133}$ | $P^{164}$ |
| | $f^{143}$ | $f^{150}$ | | | | | | | | $p^{165}$ |
| | $f^{144}$ | $f^{149}$ | | | | | | | | $p^{166}$ |
| | $f^{145}$ | $f^{148}$ | | | | | | | | $p^{167}$ |
| | $f^{146}$ | $f^{147}$ | | | | | | | | $p^{168}$ |

**Fig.10. Period 8**

| | | | | | | | | | | |
|---|---|---|---|---|---|---|---|---|---|---|
| $s^{169}$ | | | | | | | | | | |
| $s^{170}$ | $d^{171}$<br>$f^{172}$ | $d^{204}$<br>$f^{203}$ | $d^{205}$ | $d^{206}$ | $d^{207}$ | $d^{208}$ | $d^{209}$ | $d^{210}$ | $d^{211}$ | $d^{212}$ |
| | $f^{173}$<br>$g^{174}$ | $f^{202}$<br>$g^{175}$ | $g^{176}$ | $g^{177}$ | $g^{178}$ | $g^{179}$ | $g^{180}$ | $g^{181}$ | $g^{182}$ | $P^{213}$ |
| | $f^{192}$<br>$g^{191}$ | $f^{201}$<br>$g^{190}$ | $g^{189}$ | $g^{188}$ | $g^{187}$ | $g^{186}$ | $g^{185}$ | $g^{184}$ | $g^{183}$ | $P^{214}$ |
| | $f^{193}$ | $f^{200}$ | | | | | | | | $P^{215}$ |
| | $f^{194}$ | $f^{199}$ | | | | | | | | $P^{216}$ |
| | $f^{195}$ | $f^{198}$ | | | | | | | | $P^{217}$ |
| | $f^{196}$ | $f^{197}$ | | | | | | | | $P^{218}$ |

**Fig.11. Period 9**

| | | | | | | | | | | |
|---|---|---|---|---|---|---|---|---|---|---|
| $s^{219}$ | | | | | | | | | | |
| $s^{220}$ | $d^{221}$<br>$f^{222}$ | $d^{276}$<br>$f^{275}$ | $d^{277}$ | $d^{278}$ | $d^{279}$ | $d^{280}$ | $d^{281}$ | $d^{282}$ | $d^{283}$ | $d^{284}$ |
| | $f^{223}$<br>$g^{224}$ | $f^{274}$<br>$g^{225}$ | $g^{248}$<br>$h^{226}$ | $g^{249}$<br>$h^{237}$ | $g^{250}$<br>$h^{238}$ | $g^{251}$ | $g^{252}$ | $g^{253}$ | $g^{254}$ | $p^{285}$ |
| | $f^{264}$<br>$g^{263}$ | $f^{273}$<br>$g^{262}$ | $g^{261}$<br>$h^{227}$ | $g^{260}$<br>$h^{236}$ | $g^{259}$<br>$h^{239}$ | $g^{258}$ | $g^{257}$ | $g^{256}$ | $g^{255}$ | $p^{286}$ |
| | $f^{265}$ | $f^{272}$ | $h^{228}$ | $h^{235}$ | $h^{240}$ | $h^{247}$ | | | | $p^{287}$ |
| | $f^{266}$ | $f^{271}$ | $h^{229}$ | $h^{234}$ | $h^{241}$ | $h^{246}$ | | | | $p^{288}$ |
| | $f^{267}$ | $f^{270}$ | $h^{230}$ | $h^{233}$ | $h^{242}$ | $h^{245}$ | | | | $p^{289}$ |
| | $f^{268}$ | $f^{269}$ | $h^{231}$ | $h^{232}$ | $h^{243}$ | $h^{244}$ | | | | $p^{290}$ |

**Fig.12. Period 10**

| | | | | | | | | | | |
|---|---|---|---|---|---|---|---|---|---|---|
| $s^{291}$ | | | | | | | | | | |
| $s^{292}$ | $d^{293}$ $f^{294}$ | $d^{348}$ $f^{347}$ | $d^{349}$ | $d^{350}$ | $d^{351}$ | $d^{352}$ | $d^{353}$ | $d^{354}$ | $d^{355}$ | $d^{356}$ |
| | $f^{295}$ $g^{296}$ | $f^{346}$ $g^{297}$ | $g^{320}$ $h^{298}$ | $g^{321}$ $h^{309}$ | $g^{322}$ $h^{310}$ | $g^{323}$ | $g^{324}$ | $g^{325}$ | $g^{326}$ | $p^{357}$ |
| | $f^{336}$ $g^{335}$ | $f^{345}$ $g^{334}$ | $g^{333}$ $h^{299}$ | $g^{332}$ $h^{308}$ | $g^{331}$ $h^{311}$ | $g^{330}$ | $g^{329}$ | $g^{328}$ | $g^{327}$ | $p^{358}$ |
| | $f^{337}$ | $f^{344}$ | $h^{300}$ | $h^{307}$ | $h^{312}$ | $h^{319}$ | | | | $p^{359}$ |
| | $f^{338}$ | $f^{343}$ | $h^{301}$ | $h^{306}$ | $h^{313}$ | $h^{318}$ | | | | $p^{360}$ |
| | $f^{339}$ | $f^{342}$ | $h^{302}$ | $h^{305}$ | $h^{314}$ | $h^{317}$ | | | | $p^{361}$ |
| | $f^{340}$ | $f^{341}$ | $h^{303}$ | $h^{304}$ | $h^{315}$ | $h^{316}$ | | | | $p^{362}$ |

**Fig.13. Period 11**

| | | | | | | | | | | |
|---|---|---|---|---|---|---|---|---|---|---|
| $s^{363}$ | | | | | | | | | | |
| $s^{364}$ | $d^{365}$<br>$f^{366}$ | $d^{446}$<br>$f^{445}$ | $d^{447}$ | $d^{448}$ | $d^{449}$ | $d^{450}$ | $d^{451}$ | $d^{452}$ | $d^{453}$ | $d^{454}$ |
| | $f^{367}$<br>$g^{368}$ | $f^{444}$<br>$g^{369}$ | $h^{370}$<br>$g^{418}$ | $h^{407}$<br>$g^{419}$ | $h^{408}$<br>$g^{420}$ | $g^{421}$ | $g^{422}$ | $g^{423}$ | $g^{424}$ | $p^{455}$ |
| | $f^{434}$<br>$g^{433}$ | $f^{443}$<br>$g^{432}$ | $h^{371}$<br>$g^{431}$ | $h^{406}$<br>$g^{430}$ | $h^{409}$<br>$g^{429}$ | $g^{428}$ | $g^{427}$ | $g^{426}$ | $g^{425}$ | $p^{456}$ |
| | $f^{435}$ | $f^{442}$ | $h^{372}$<br>$i^{373}$ | $h^{405}$<br>$i^{374}$ | $h^{410}$<br>$i^{375}$ | $h^{417}$<br>$i^{376}$ | $i^{377}$ | $i^{378}$ | $i^{379}$ | $p^{457}$ |
| | $f^{436}$ | $f^{441}$ | $h^{399}$<br>$i^{386}$ | $h^{404}$<br>$i^{385}$ | $h^{411}$<br>$i^{384}$ | $h^{416}$<br>$i^{383}$ | $i^{382}$ | $i^{381}$ | $i^{380}$ | $p^{458}$ |
| | $f^{437}$ | $f^{440}$ | $h^{400}$<br>$i^{387}$ | $h^{403}$<br>$i^{388}$ | $h^{412}$<br>$i^{389}$ | $h^{415}$<br>$i^{390}$ | $i^{391}$ | $i^{392}$ | $i^{393}$ | $p^{459}$ |
| | $f^{438}$ | $f^{439}$ | $h^{401}$ | $h^{402}$ | $h^{413}$<br>$i^{398}$ | $h^{414}$<br>$i^{397}$ | $i^{396}$ | $i^{395}$ | $i^{394}$ | $p^{460}$ |

**Fig.14. Period 12**

| | | | | | | | | | | |
|---|---|---|---|---|---|---|---|---|---|---|
| $s^{461}$ | | | | | | | | | | |
| $s^{462}$ | $d^{463}$<br>$f^{464}$ | $d^{544}$<br>$f^{543}$ | $d^{545}$ | $d^{546}$ | $d^{547}$ | $d^{548}$ | $d^{549}$ | $d^{550}$ | $d^{551}$ | $d^{552}$ |
| | $f^{465}$<br>$g^{466}$ | $f^{542}$<br>$g^{467}$ | $h^{468}$<br>$g^{516}$ | $h^{505}$<br>$g^{517}$ | $h^{506}$<br>$g^{518}$ | $g^{519}$ | $g^{520}$ | $g^{521}$ | $g^{522}$ | $p^{553}$ |
| | $f^{532}$<br>$g^{531}$ | $f^{541}$<br>$g^{530}$ | $h^{469}$<br>$g^{529}$ | $h^{504}$<br>$g^{528}$ | $h^{507}$<br>$g^{527}$ | $g^{526}$ | $g^{525}$ | $g^{524}$ | $g^{523}$ | $p^{554}$ |
| | $f^{533}$ | $f^{540}$ | $h^{470}$<br>$i^{471}$ | $h^{503}$<br>$i^{472}$ | $h^{508}$<br>$i^{473}$ | $h^{515}$<br>$i^{474}$ | $i^{475}$ | $i^{476}$ | $i^{477}$ | $p^{555}$ |
| | $f^{534}$ | $f^{539}$ | $h^{497}$<br>$i^{484}$ | $h^{502}$<br>$i^{483}$ | $h^{509}$<br>$i^{482}$ | $h^{514}$<br>$i^{481}$ | $i^{480}$ | $i^{479}$ | $i^{478}$ | $p^{556}$ |
| | $f^{535}$ | $f^{538}$ | $h^{498}$<br>$i^{485}$ | $h^{501}$<br>$i^{486}$ | $h^{510}$<br>$i^{487}$ | $h^{513}$<br>$i^{488}$ | $i^{489}$ | $i^{490}$ | $i^{491}$ | $p^{557}$ |
| | $f^{536}$ | $f^{537}$ | $h^{499}$ | $h^{500}$ | $h^{511}$<br>$i^{496}$ | $h^{512}$<br>$i^{495}$ | $i^{494}$ | $i^{493}$ | $i^{492}$ | $p^{558}$ |

**Fig.15. Period 13**

| | | | | | | | | | | |
|---|---|---|---|---|---|---|---|---|---|---|
| $s^{559}$ | | | | | | | | | | |
| $s^{560}$ | $d^{561}$<br>$f^{562}$ | $d^{672}$<br>$f^{671}$ | $d^{673}$ | $d^{674}$ | $d^{675}$ | $d^{676}$ | $d^{677}$ | $d^{678}$ | $d^{679}$ | $d^{680}$ |
| | $f^{563}$<br>$g^{564}$ | $f^{670}$<br>$g^{565}$ | $h^{566}$<br>$g^{644}$ | $h^{633}$<br>$g^{645}$ | $h^{634}$<br>$g^{646}$ | $g^{647}$ | $g^{648}$ | $g^{649}$ | $g^{650}$ | $p^{681}$ |
| | $f^{660}$<br>$g^{659}$ | $f^{669}$<br>$g^{658}$ | $h^{567}$<br>$g^{657}$ | $h^{632}$<br>$g^{656}$ | $h^{635}$<br>$g^{655}$ | $g^{654}$ | $g^{653}$ | $g^{652}$ | $g^{651}$ | $p^{682}$ |
| | $f^{661}$<br>$j^{569}$ | $f^{668}$<br>$j^{570}$ | $h^{624}$<br>$i^{568}$<br>$j^{571}$ | $h^{631}$<br>$i^{599}$<br>$j^{572}$ | $h^{636}$<br>$i^{600}$<br>$j^{573}$ | $h^{643}$<br>$i^{601}$<br>$j^{574}$ | $i^{602}$<br>$j^{575}$ | $i^{603}$<br>$j^{576}$ | $i^{604}$<br>$j^{577}$ | $p^{683}$ |
| | $f^{662}$<br>$j^{586}$ | $f^{667}$<br>$j^{585}$ | $h^{625}$<br>$i^{611}$<br>$j^{584}$ | $h^{630}$<br>$i^{610}$<br>$j^{583}$ | $h^{637}$<br>$i^{609}$<br>$j^{582}$ | $h^{642}$<br>$i^{608}$<br>$j^{581}$ | $i^{607}$<br>$j^{580}$ | $i^{606}$<br>$j^{579}$ | $i^{605}$<br>$j^{578}$ | $p^{684}$ |
| | $f^{663}$<br>$j^{587}$ | $f^{666}$<br>$j^{588}$ | $h^{626}$<br>$i^{612}$<br>$j^{589}$ | $h^{629}$<br>$i^{613}$<br>$j^{590}$ | $h^{638}$<br>$i^{614}$<br>$j^{591}$ | $h^{641}$<br>$i^{615}$<br>$j^{592}$ | $i^{616}$<br>$j^{593}$ | $i^{617}$<br>$j^{594}$ | $i^{618}$<br>$j^{595}$ | $p^{685}$ |
| | $f^{664}$ | $f^{665}$ | $h^{627}$ | $h^{628}$ | $h^{639}$<br>$i^{623}$ | $h^{640}$<br>$i^{622}$ | $i^{621}$<br>$j^{598}$ | $i^{620}$<br>$j^{597}$ | $i^{619}$<br>$j^{596}$ | $p^{686}$ |

**Fig.16. Period 14**

| | | | | | | | | | | |
|---|---|---|---|---|---|---|---|---|---|---|
| $s^{687}$ | | | | | | | | | | |
| $s^{688}$ | $d^{689}$<br>$f^{690}$ | $d^{800}$<br>$f^{799}$ | $d^{801}$ | $d^{802}$ | $d^{803}$ | $d^{804}$ | $d^{805}$ | $d^{806}$ | $d^{807}$ | $d^{808}$ |
| | $f^{691}$<br>$g^{692}$ | $f^{798}$<br>$g^{693}$ | $h^{694}$<br>$g^{772}$ | $h^{761}$<br>$g^{773}$ | $h^{762}$<br>$g^{774}$ | $g^{775}$ | $g^{776}$ | $g^{777}$ | $g^{778}$ | $p^{809}$ |
| | $f^{788}$<br>$g^{787}$ | $f^{797}$<br>$g^{786}$ | $h^{695}$<br>$g^{785}$ | $h^{760}$<br>$g^{784}$ | $h^{763}$<br>$g^{783}$ | $g^{782}$ | $g^{781}$ | $g^{780}$ | $g^{779}$ | $p^{810}$ |
| | $f^{789}$<br>$j^{697}$ | $f^{796}$<br>$j^{698}$ | $h^{752}$<br>$i^{696}$<br>$j^{699}$ | $h^{759}$<br>$i^{727}$<br>$j^{700}$ | $h^{764}$<br>$i^{728}$<br>$j^{701}$ | $h^{771}$<br>$i^{729}$<br>$j^{702}$ | $i^{730}$<br>$j^{703}$ | $i^{731}$<br>$j^{704}$ | $i^{732}$<br>$j^{705}$ | $p^{811}$ |
| | $f^{790}$<br>$j^{714}$ | $f^{795}$<br>$j^{713}$ | $h^{753}$<br>$i^{739}$<br>$j^{712}$ | $h^{758}$<br>$i^{738}$<br>$j^{711}$ | $h^{765}$<br>$i^{737}$<br>$j^{710}$ | $h^{770}$<br>$i^{736}$<br>$j^{709}$ | $i^{735}$<br>$j^{708}$ | $i^{734}$<br>$j^{707}$ | $i^{733}$<br>$j^{706}$ | $p^{812}$ |
| | $f^{791}$<br>$j^{715}$ | $f^{794}$<br>$j^{716}$ | $h^{754}$<br>$i^{740}$<br>$j^{717}$ | $h^{757}$<br>$i^{741}$<br>$j^{718}$ | $h^{766}$<br>$i^{742}$<br>$j^{719}$ | $h^{769}$<br>$i^{743}$<br>$j^{720}$ | $i^{744}$<br>$j^{721}$ | $i^{745}$<br>$j^{722}$ | $i^{746}$<br>$j^{723}$ | $p^{813}$ |
| | $f^{792}$ | $f^{793}$ | $h^{755}$ | $h^{756}$ | $h^{767}$<br>$i^{751}$ | $h^{768}$<br>$i^{750}$ | $i^{749}$<br>$j^{726}$ | $i^{748}$<br>$j^{725}$ | $i^{747}$<br>$j^{724}$ | $p^{814}$ |

**Fig.17. Period 15**

| | | | | | | | | | | |
|---|---|---|---|---|---|---|---|---|---|---|
| $s^{815}$ | | | | | | | | | | |
| $s^{816}$ | $d^{817}$<br>$f^{818}$ | $d^{962}$<br>$f^{961}$ | $d^{963}$ | $d^{964}$ | $d^{965}$ | $d^{966}$ | $d^{967}$ | $d^{968}$ | $d^{969}$ | $d^{970}$ |
| | $f^{819}$<br>$g^{820}$ | $f^{960}$<br>$g^{821}$ | $h^{822}$<br>$g^{934}$ | $h^{923}$<br>$g^{935}$ | $h^{924}$<br>$g^{936}$ | $g^{937}$<br>$k^{827}$ | $g^{938}$<br>$k^{828}$ | $g^{939}$<br>$k^{829}$ | $g^{940}$<br>$k^{830}$ | $p^{971}$ |
| | $f^{950}$<br>$g^{949}$ | $f^{959}$<br>$g^{948}$ | $h^{823}$<br>$g^{947}$ | $h^{922}$<br>$g^{946}$ | $h^{925}$<br>$g^{945}$ | $g^{944}$<br>$k^{834}$ | $g^{943}$<br>$k^{833}$ | $g^{942}$<br>$k^{832}$ | $g^{941}$<br>$k^{831}$ | $p^{972}$ |
| | $f^{951}$<br>$j^{825}$ | $f^{958}$<br>$j^{826}$ | $h^{914}$<br>$i^{824}$<br>$j^{861}$ | $h^{921}$<br>$i^{889}$<br>$j^{862}$ | $h^{926}$<br>$i^{890}$<br>$j^{863}$ | $h^{933}$<br>$i^{891}$<br>$j^{864}$<br>$k^{835}$ | $i^{892}$<br>$j^{865}$<br>$k^{836}$ | $i^{893}$<br>$j^{866}$<br>$k^{837}$ | $i^{894}$<br>$j^{867}$<br>$k^{838}$ | $p^{973}$ |
| | $f^{952}$<br>$j^{876}$ | $f^{957}$<br>$j^{875}$ | $h^{915}$<br>$i^{901}$<br>$j^{874}$ | $h^{920}$<br>$i^{900}$<br>$j^{873}$ | $h^{927}$<br>$i^{899}$<br>$j^{872}$ | $h^{932}$<br>$i^{898}$<br>$j^{871}$<br>$k^{642}$ | $i^{897}$<br>$j^{870}$<br>$k^{841}$ | $i^{896}$<br>$j^{869}$<br>$k^{840}$ | $i^{895}$<br>$j^{868}$<br>$k^{839}$ | $p^{974}$ |
| | $f^{953}$<br>$j^{877}$<br>$k^{843}$ | $f^{956}$<br>$j^{878}$<br>$k^{844}$ | $h^{916}$<br>$i^{902}$<br>$j^{879}$<br>$k^{845}$ | $h^{919}$<br>$i^{903}$<br>$j^{880}$<br>$k^{846}$ | $h^{928}$<br>$i^{904}$<br>$j^{881}$<br>$k^{847}$ | $h^{931}$<br>$i^{905}$<br>$j^{882}$<br>$k^{848}$ | $i^{906}$<br>$j^{883}$<br>$k^{849}$ | $i^{907}$<br>$j^{884}$<br>$k^{850}$ | $i^{908}$<br>$j^{885}$<br>$k^{851}$ | $p^{975}$ |
| | $f^{954}$<br>$k^{860}$ | $f^{955}$<br>$k^{859}$ | $h^{917}$<br>$k^{858}$ | $h^{918}$<br>$k^{857}$ | $h^{929}$<br>$i^{913}$<br>$k^{856}$ | $h^{930}$<br>$i^{912}$<br>$k^{855}$ | $i^{911}$<br>$j^{888}$<br>$k^{854}$ | $i^{910}$<br>$j^{887}$<br>$k^{853}$ | $i^{909}$<br>$j^{886}$<br>$k^{852}$ | $p^{976}$ |

**Fig.18. Period 16**

| | | | | | | | | | | |
|---|---|---|---|---|---|---|---|---|---|---|
| $s^{977}$ | | | | | | | | | | |
| $s^{978}$ | $d^{979}$<br>$f^{980}$ | $d^{1124}$<br>$f^{1123}$ | $d^{1125}$ | $d^{1126}$ | $d^{1127}$ | $d^{1128}$ | $d^{1129}$ | $d^{1130}$ | $d^{1131}$ | $d^{1132}$ |
| | $f^{981}$<br>$g^{982}$ | $f^{1122}$<br>$g^{983}$ | $h^{984}$<br>$g^{1096}$ | $h^{1085}$<br>$g^{1097}$ | $h^{1086}$<br>$g^{1098}$ | $g^{1099}$<br>$k^{989}$ | $g^{1100}$<br>$k^{990}$ | $g^{1101}$<br>$k^{991}$ | $g^{1102}$<br>$k^{992}$ | $p^{1133}$ |
| | $f^{1112}$<br>$g^{1111}$ | $f^{1121}$<br>$g^{1110}$ | $h^{985}$<br>$g^{1109}$ | $h^{1084}$<br>$g^{1108}$ | $h^{1087}$<br>$g^{1107}$ | $g^{1106}$<br>$k^{996}$ | $g^{1105}$<br>$k^{995}$ | $g^{1104}$<br>$k^{994}$ | $g^{1103}$<br>$k^{993}$ | $p^{1134}$ |
| | $f^{1113}$<br>$j^{987}$ | $f^{1120}$<br>$j^{988}$ | $h^{1076}$<br>$i^{986}$<br>$j^{1023}$ | $h^{1083}$<br>$i^{1051}$<br>$j^{1024}$ | $h^{1088}$<br>$i^{1052}$<br>$j^{1025}$ | $h^{1095}$<br>$i^{1053}$<br>$j^{1026}$<br>$k^{997}$ | $i^{1054}$<br>$j^{1027}$<br>$k^{998}$ | $i^{1055}$<br>$j^{1028}$<br>$k^{999}$ | $i^{1056}$<br>$j^{1029}$<br>$k^{1000}$ | $p^{1135}$ |
| | $f^{1114}$<br>$j^{1038}$ | $f^{1119}$<br>$j^{1037}$ | $h^{1077}$<br>$i^{1063}$<br>$j^{1036}$ | $h^{1082}$<br>$i^{1062}$<br>$j^{1035}$ | $h^{1089}$<br>$i^{1061}$<br>$j^{1034}$ | $h^{1094}$<br>$i^{1060}$<br>$j^{1033}$<br>$k^{1004}$ | $i^{1059}$<br>$j^{1032}$<br>$k^{1003}$ | $i^{1058}$<br>$j^{1031}$<br>$k^{1002}$ | $i^{1057}$<br>$j^{1030}$<br>$k^{1001}$ | $p^{1136}$ |
| | $f^{1115}$<br>$j^{1039}$<br>$k^{1005}$ | $f^{1118}$<br>$j^{1040}$<br>$k^{1006}$ | $h^{1078}$<br>$i^{1064}$<br>$j^{1041}$<br>$k^{1007}$ | $h^{1081}$<br>$i^{1065}$<br>$j^{1042}$<br>$k^{1008}$ | $h^{1090}$<br>$i^{1066}$<br>$j^{1043}$<br>$k^{1009}$ | $h^{1093}$<br>$i^{1067}$<br>$j^{1044}$<br>$k^{1010}$ | $i^{1068}$<br>$j^{1045}$<br>$k^{1011}$ | $i^{1069}$<br>$j^{1046}$<br>$k^{1012}$ | $i^{1070}$<br>$j^{1047}$<br>$k^{1013}$ | $p^{1137}$ |
| | $f^{1116}$<br>$k^{1022}$ | $f^{1117}$<br>$k^{1021}$ | $h^{1079}$<br>$k^{1020}$ | $h^{1080}$<br>$k^{1019}$ | $h^{1091}$<br>$i^{1075}$<br>$k^{1018}$ | $h^{1092}$<br>$i^{1074}$<br>$k^{1017}$ | $i^{1073}$<br>$j^{1050}$<br>$k^{1016}$ | $i^{1072}$<br>$j^{1049}$<br>$k^{1015}$ | $i^{1071}$<br>$j^{1048}$<br>$k^{1014}$ | $p^{1138}$ |

**Fig.19. Period 17**

| | | | | | | | | | | |
|---|---|---|---|---|---|---|---|---|---|---|
| $s^{1139}$ | | | | | | | | | | |
| $s^{1140}$ | $d^{1141}$<br>$f^{1142}$ | $d^{1324}$<br>$f^{1323}$ | $d^{1325}$ | $d^{1326}$ | $d^{1327}$ | $d^{1328}$ | $d^{1329}$ | $d^{1330}$ | $d^{1331}$ | $d^{1332}$ |
| | $f^{1143}$<br>$g^{1144}$<br>$v^{1152}$ | $f^{1322}$<br>$g^{1145}$<br>$v^{1153}$ | $h^{1146}$<br>$g^{1296}$<br>$v^{1154}$ | $h^{1285}$<br>$g^{1297}$<br>$v^{1155}$ | $h^{1286}$<br>$g^{1298}$<br>$v^{1156}$ | $g^{1299}$<br>$k^{1151}$<br>$v^{1157}$ | $g^{1300}$<br>$k^{1190}$<br>$v^{1158}$ | $g^{1301}$<br>$k^{1191}$<br>$v^{1159}$ | $g^{1302}$<br>$k^{1192}$<br>$v^{1160}$ | $p^{1333}$ |
| | $f^{1312}$<br>$g^{1311}$<br>$v^{1169}$ | $f^{1321}$<br>$g^{1310}$<br>$v^{1168}$ | $h^{1147}$<br>$g^{1309}$<br>$v^{1167}$ | $h^{1284}$<br>$g^{1308}$<br>$v^{1166}$ | $h^{1287}$<br>$g^{1307}$<br>$v^{1165}$ | $g^{1306}$<br>$k^{1196}$<br>$v^{1164}$ | $g^{1305}$<br>$k^{1195}$<br>$v^{1163}$ | $g^{1304}$<br>$k^{1194}$<br>$v^{1162}$ | $g^{1303}$<br>$k^{1193}$<br>$v^{1161}$ | $p^{1334}$ |
| | $f^{1313}$<br>$j^{1149}$<br>$v^{1170}$ | $f^{1320}$<br>$j^{1150}$<br>$v^{1171}$ | $h^{1276}$<br>$i^{1148}$<br>$j^{1223}$<br>$v^{1172}$ | $h^{1283}$<br>$i^{1251}$<br>$j^{1224}$<br>$v^{1173}$ | $h^{1288}$<br>$i^{1252}$<br>$j^{1225}$<br>$v^{1174}$ | $h^{1295}$<br>$i^{1253}$<br>$j^{1226}$<br>$k^{1197}$<br>$v^{1175}$ | $i^{1254}$<br>$j^{1227}$<br>$k^{1198}$<br>$v^{1176}$ | $i^{1255}$<br>$j^{1228}$<br>$k^{1199}$<br>$v^{1177}$ | $i^{1256}$<br>$j^{1229}$<br>$k^{1200}$<br>$v^{1178}$ | $p^{1335}$ |
| | $f^{1314}$<br>$j^{1238}$<br>$v^{1187}$ | $f^{1319}$<br>$j^{1237}$<br>$v^{1186}$ | $h^{1277}$<br>$i^{1263}$<br>$j^{1236}$<br>$v^{1185}$ | $h^{1282}$<br>$i^{1262}$<br>$j^{1235}$<br>$v^{1184}$ | $h^{1289}$<br>$i^{1261}$<br>$j^{1234}$<br>$v^{1183}$ | $h^{1294}$<br>$i^{1260}$<br>$j^{1233}$<br>$k^{1204}$<br>$v^{1182}$ | $i^{1259}$<br>$j^{1232}$<br>$k^{1203}$<br>$v^{1181}$ | $i^{1258}$<br>$j^{1231}$<br>$k^{1202}$<br>$v^{1180}$ | $i^{1257}$<br>$j^{1230}$<br>$k^{1201}$<br>$v^{1179}$ | $p^{1336}$ |
| | $f^{1315}$<br>$j^{1239}$<br>$k^{1205}$<br>$v^{1188}$ | $f^{1318}$<br>$j^{1240}$<br>$k^{1206}$<br>$v^{1189}$ | $h^{1278}$<br>$i^{1264}$<br>$j^{1241}$<br>$k^{1207}$ | $h^{1281}$<br>$i^{1265}$<br>$j^{1242}$<br>$k^{1208}$ | $h^{1290}$<br>$i^{1266}$<br>$j^{1243}$<br>$k^{1209}$ | $h^{1293}$<br>$i^{1267}$<br>$j^{1244}$<br>$k^{1210}$ | $i^{1268}$<br>$j^{1245}$<br>$k^{1211}$ | $i^{1269}$<br>$j^{1246}$<br>$k^{1212}$ | $i^{1270}$<br>$j^{1247}$<br>$k^{1213}$ | $p^{1337}$ |
| | $f^{1316}$<br>$k^{1222}$ | $f^{1317}$<br>$k^{1221}$ | $h^{1279}$<br>$k^{1220}$ | $h^{1280}$<br>$k^{1219}$ | $h^{1291}$<br>$i^{1275}$<br>$k^{1218}$ | $h^{1292}$<br>$i^{1274}$<br>$k^{1217}$ | $i^{1273}$<br>$j^{1250}$<br>$k^{1216}$ | $i^{1272}$<br>$j^{1249}$<br>$k^{1215}$ | $i^{1271}$<br>$j^{1248}$<br>$k^{1214}$ | $p^{1338}$ |

**Fig.20. Period 18**

| | | | | | | | | | | |
|---|---|---|---|---|---|---|---|---|---|---|
| $s^{1339}$ | | | | | | | | | | |
| $s^{1340}$ | $d^{1341}$<br>$f^{1342}$ | $d^{1524}$<br>$f^{1523}$ | $d^{1525}$ | $d^{1526}$ | $d^{1527}$ | $d^{1528}$ | $d^{1529}$ | $d^{1530}$ | $d^{1531}$ | $d^{1532}$ |
| | $f^{1343}$<br>$g^{1344}$<br>$v^{1352}$ | $f^{1522}$<br>$g^{1345}$<br>$v^{1353}$ | $h^{1346}$<br>$g^{1496}$<br>$v^{1354}$ | $h^{1485}$<br>$g^{1497}$<br>$v^{1355}$ | $h^{1486}$<br>$g^{1498}$<br>$v^{1356}$ | $g^{1499}$<br>$k^{1351}$<br>$v^{1357}$ | $g^{1500}$<br>$k^{1390}$<br>$v^{1358}$ | $g^{1501}$<br>$k^{1391}$<br>$v^{1359}$ | $g^{1502}$<br>$k^{1392}$<br>$v^{1360}$ | $p^{1533}$ |
| | $f^{1512}$<br>$g^{1511}$<br>$v^{1369}$ | $f^{1521}$<br>$g^{1510}$<br>$v^{1368}$ | $h^{1347}$<br>$g^{1509}$<br>$v^{1367}$ | $h^{1484}$<br>$g^{1508}$<br>$v^{1366}$ | $h^{1487}$<br>$g^{1507}$<br>$v^{1365}$ | $g^{1506}$<br>$k^{1396}$<br>$v^{1364}$ | $g^{1505}$<br>$k^{1395}$<br>$v^{1363}$ | $g^{1504}$<br>$k^{1394}$<br>$v^{1362}$ | $g^{1503}$<br>$k^{1393}$<br>$v^{1361}$ | $p^{1534}$ |
| | $f^{1513}$<br>$j^{1349}$<br>$v^{1370}$ | $f^{1520}$<br>$j^{1350}$<br>$v^{1371}$ | $h^{1476}$<br>$i^{1348}$<br>$j^{1423}$<br>$v^{1372}$ | $h^{1483}$<br>$i^{1451}$<br>$j^{1424}$<br>$v^{1373}$ | $h^{1488}$<br>$i^{1452}$<br>$j^{1425}$<br>$v^{1374}$ | $h^{1495}$<br>$i^{1453}$<br>$j^{1426}$<br>$k^{1397}$<br>$v^{1375}$ | $i^{1454}$<br>$j^{1427}$<br>$k^{1398}$<br>$v^{1376}$ | $i^{1455}$<br>$j^{1428}$<br>$k^{1399}$<br>$v^{1377}$ | $i^{1456}$<br>$j^{1429}$<br>$k^{1400}$<br>$v^{1378}$ | $p^{1535}$ |
| | $f^{1514}$<br>$j^{1438}$<br>$v^{1387}$ | $f^{1519}$<br>$j^{1437}$<br>$v^{1386}$ | $h^{1477}$<br>$i^{1463}$<br>$j^{1436}$<br>$v^{1385}$ | $h^{1482}$<br>$i^{1462}$<br>$j^{1435}$<br>$v^{1384}$ | $h^{1489}$<br>$i^{1461}$<br>$j^{1434}$<br>$v^{1383}$ | $h^{1494}$<br>$i^{1460}$<br>$j^{1433}$<br>$k^{1404}$<br>$v^{1382}$ | $i^{1459}$<br>$j^{1432}$<br>$k^{1403}$<br>$v^{1381}$ | $i^{1458}$<br>$j^{1431}$<br>$k^{1402}$<br>$v^{1380}$ | $i^{1457}$<br>$j^{1430}$<br>$k^{1401}$<br>$v^{1379}$ | $p^{1536}$ |
| | $f^{1515}$<br>$j^{1439}$<br>$k^{1405}$<br>$v^{1388}$ | $f^{1518}$<br>$j^{1440}$<br>$k^{1406}$<br>$v^{1389}$ | $h^{1478}$<br>$i^{1464}$<br>$j^{1441}$<br>$k^{1407}$ | $h^{1481}$<br>$i^{1465}$<br>$j^{1442}$<br>$k^{1408}$ | $h^{1490}$<br>$i^{1466}$<br>$j^{1443}$<br>$k^{1409}$ | $h^{1493}$<br>$i^{1467}$<br>$j^{1444}$<br>$k^{1410}$ | $i^{1468}$<br>$j^{1445}$<br>$k^{1411}$ | $i^{1469}$<br>$j^{1446}$<br>$k^{1412}$ | $i^{1470}$<br>$j^{1447}$<br>$k^{1413}$ | $p^{1537}$ |
| | $f^{1516}$<br>$k^{1422}$ | $f^{1517}$<br>$k^{1421}$ | $h^{1479}$<br>$k^{1420}$ | $h^{1480}$<br>$k^{1419}$ | $h^{1491}$<br>$i^{1475}$<br>$k^{1418}$ | $h^{1492}$<br>$i^{1474}$<br>$k^{1417}$ | $i^{1473}$<br>$j^{1450}$<br>$k^{1416}$ | $i^{1472}$<br>$j^{1449}$<br>$k^{1415}$ | $i^{1471}$<br>$j^{1448}$<br>$k^{1414}$ | $p^{1538}$ |

**Fig.21. Period 19**

| | | | | | | | | | | |
|---|---|---|---|---|---|---|---|---|---|---|
| $s^{1539}$ | | | | | | | | | | |
| $s^{1540}$ | $d^{1541}$<br>$f^{1542}$ | $d^{1766}$<br>$f^{1765}$ | $d^{1767}$ | $d^{1768}$ | $d^{1769}$ | $d^{1770}$ | $d^{1771}$ | $d^{1772}$ | $d^{1773}$ | $d^{1774}$ |
| | $f^{1543}$<br>$g^{1544}$<br>$v^{1552}$ | $f^{1764}$<br>$g^{1545}$<br>$v^{1553}$ | $h^{1546}$<br>$g^{1738}$<br>$v^{1596}$<br>$w^{1554}$ | $h^{1727}$<br>$g^{1739}$<br>$v^{1597}$<br>$w^{1555}$ | $h^{1728}$<br>$g^{1740}$<br>$v^{1598}$<br>$w^{1556}$ | $g^{1741}$<br>$k^{1551}$<br>$v^{1599}$<br>$w^{1557}$ | $g^{1742}$<br>$k^{1632}$<br>$v^{1600}$<br>$w^{1558}$ | $g^{1743}$<br>$k^{1633}$<br>$v^{1601}$<br>$w^{1559}$ | $g^{1744}$<br>$k^{1634}$<br>$v^{1602}$<br>$w^{1560}$ | $p^{1775}$ |
| | $f^{1754}$<br>$g^{1753}$<br>$v^{1611}$ | $f^{1763}$<br>$g^{1752}$<br>$v^{1610}$ | $h^{1547}$<br>$g^{1751}$<br>$v^{1609}$<br>$w^{1567}$ | $h^{1726}$<br>$g^{1750}$<br>$v^{1608}$<br>$w^{1566}$ | $h^{1729}$<br>$g^{1749}$<br>$v^{1607}$<br>$w^{1565}$ | $g^{1748}$<br>$k^{1638}$<br>$v^{1606}$<br>$w^{1564}$ | $g^{1747}$<br>$k^{1637}$<br>$v^{1605}$<br>$w^{1563}$ | $g^{1746}$<br>$k^{1636}$<br>$v^{1604}$<br>$w^{1562}$ | $g^{1745}$<br>$k^{1635}$<br>$v^{1603}$<br>$w^{1561}$ | $p^{1776}$ |
| | $f^{1755}$<br>$j^{1549}$<br>$v^{1612}$ | $f^{1762}$<br>$j^{1550}$<br>$v^{1613}$ | $h^{1718}$<br>$w^{1568}$<br>$i^{1548}$<br>$j^{1665}$<br>$v^{1614}$ | $h^{1725}$<br>$w^{1569}$<br>$i^{1693}$<br>$j^{1666}$<br>$v^{1615}$ | $h^{1730}$<br>$w^{1570}$<br>$i^{1694}$<br>$j^{1667}$<br>$v^{1616}$ | $h^{1737}$<br>$w^{1571}$<br>$i^{1695}$<br>$j^{1668}$<br>$k^{1639}$<br>$v^{1617}$ | $i^{1696}$<br>$w^{1572}$<br>$j^{1669}$<br>$k^{1640}$<br>$v^{1618}$ | $i^{1697}$<br>$w^{1573}$<br>$j^{1670}$<br>$k^{1641}$<br>$v^{1619}$ | $i^{1698}$<br>$w^{1574}$<br>$j^{1671}$<br>$k^{1642}$<br>$v^{1620}$ | $p^{1777}$ |
| | $f^{1756}$<br>$j^{1680}$<br>$v^{1629}$ | $f^{1761}$<br>$j^{1679}$<br>$v^{1628}$ | $h^{1719}$<br>$w^{1581}$<br>$i^{1705}$<br>$j^{1678}$<br>$v^{1627}$ | $h^{1724}$<br>$w^{1580}$<br>$i^{1704}$<br>$j^{1677}$<br>$v^{1626}$ | $h^{1731}$<br>$w^{1579}$<br>$i^{1703}$<br>$j^{1676}$<br>$v^{1625}$ | $h^{1736}$<br>$w^{1578}$<br>$i^{1702}$<br>$j^{1675}$<br>$k^{1646}$<br>$v^{1624}$ | $i^{1701}$<br>$w^{1577}$<br>$j^{1674}$<br>$k^{1645}$<br>$v^{1623}$ | $i^{1700}$<br>$w^{1576}$<br>$j^{1673}$<br>$k^{1644}$<br>$v^{1622}$ | $i^{1699}$<br>$w^{1575}$<br>$j^{1672}$<br>$k^{1643}$<br>$v^{1621}$ | $p^{1778}$ |
| | $f^{1757}$<br>$j^{1681}$<br>$k^{1647}$<br>$v^{1630}$ | $f^{1760}$<br>$j^{1682}$<br>$k^{1648}$<br>$v^{1631}$ | $h^{1720}$<br>$w^{1582}$<br>$i^{1706}$<br>$j^{1683}$<br>$k^{1649}$ | $h^{1723}$<br>$w^{1583}$<br>$i^{1707}$<br>$j^{1684}$<br>$k^{1650}$ | $h^{1732}$<br>$w^{1584}$<br>$i^{1708}$<br>$j^{1685}$<br>$k^{1651}$ | $h^{1735}$<br>$w^{1585}$<br>$i^{1709}$<br>$j^{1686}$<br>$k^{1652}$ | $i^{1710}$<br>$w^{1586}$<br>$j^{1687}$<br>$k^{1653}$ | $i^{1711}$<br>$w^{1587}$<br>$j^{1688}$<br>$k^{1654}$ | $i^{1712}$<br>$w^{1588}$<br>$j^{1689}$<br>$k^{1655}$ | $p^{1779}$ |
| | $f^{1758}$<br>$k^{1664}$ | $f^{1759}$<br>$k^{1663}$ | $h^{1721}$<br>$k^{1662}$<br>$w^{1595}$ | $h^{1722}$<br>$k^{1661}$<br>$w^{1594}$ | $h^{1733}$<br>$i^{1717}$<br>$k^{1660}$<br>$w^{1593}$ | $h^{1734}$<br>$i^{1716}$<br>$k^{1659}$<br>$w^{1592}$ | $i^{1715}$<br>$j^{1692}$<br>$k^{1658}$<br>$w^{1591}$ | $i^{1714}$<br>$j^{1691}$<br>$k^{1657}$<br>$w^{1590}$ | $i^{1713}$<br>$j^{1690}$<br>$k^{1656}$<br>$w^{1589}$ | $p^{1780}$ |

**Fig.22. Period 20**

| | | | | | | | | | | |
|---|---|---|---|---|---|---|---|---|---|---|
| $s^{1781}$ | | | | | | | | | | |
| $s^{1782}$ | $d^{1783}$<br>$f^{1784}$ | $d^{2008}$<br>$f^{2007}$ | $d^{2009}$ | $d^{2010}$ | $d^{2011}$ | $d^{2012}$ | $d^{2013}$ | $d^{2014}$ | $d^{2015}$ | $d^{2016}$ |
| | $f^{1785}$<br>$g^{1786}$<br>$v^{1794}$ | $f^{2006}$<br>$g^{1787}$<br>$v^{1795}$ | $h^{1788}$<br>$g^{1980}$<br>$v^{1838}$<br>$w^{1796}$ | $h^{1969}$<br>$g^{1981}$<br>$v^{1839}$<br>$w^{1797}$ | $h^{1970}$<br>$g^{1982}$<br>$v^{1840}$<br>$w^{1798}$ | $g^{1983}$<br>$k^{1793}$<br>$v^{1841}$<br>$w^{1799}$ | $g^{1984}$<br>$k^{1874}$<br>$v^{1842}$<br>$w^{1800}$ | $g^{1985}$<br>$k^{1875}$<br>$v^{1843}$<br>$w^{1801}$ | $g^{1986}$<br>$k^{1876}$<br>$v^{1844}$<br>$w^{1802}$ | $p^{2017}$ |
| | $f^{1996}$<br>$g^{1995}$<br>$v^{1853}$ | $f^{2005}$<br>$g^{1994}$<br>$v^{1852}$ | $h^{1789}$<br>$g^{1993}$<br>$v^{1851}$<br>$w^{1809}$ | $h^{1968}$<br>$g^{1992}$<br>$v^{1850}$<br>$w^{1808}$ | $h^{1971}$<br>$g^{1991}$<br>$v^{1849}$<br>$w^{1807}$ | $g^{1990}$<br>$k^{1880}$<br>$v^{1848}$<br>$w^{1806}$ | $g^{1989}$<br>$k^{1879}$<br>$v^{1847}$<br>$w^{1805}$ | $g^{1988}$<br>$k^{1878}$<br>$v^{1846}$<br>$w^{1804}$ | $g^{1987}$<br>$k^{1877}$<br>$v^{1845}$<br>$w^{1803}$ | $p^{2018}$ |
| | $f^{1997}$<br>$j^{1791}$<br>$v^{1854}$ | $f^{2004}$<br>$j^{1792}$<br>$v^{1855}$ | $h^{1960}$<br>$w^{1810}$<br>$i^{1790}$<br>$j^{1907}$<br>$v^{1856}$ | $h^{1967}$<br>$w^{1811}$<br>$i^{1935}$<br>$j^{1908}$<br>$v^{1857}$ | $h^{1972}$<br>$w^{1812}$<br>$i^{1936}$<br>$j^{1909}$<br>$v^{1858}$ | $h^{1979}$<br>$w^{1813}$<br>$i^{1937}$<br>$j^{1910}$<br>$k^{1881}$<br>$v^{1859}$ | $i^{1938}$<br>$w^{1814}$<br>$j^{1911}$<br>$k^{1882}$<br>$v^{1860}$ | $i^{1939}$<br>$w^{1815}$<br>$j^{1912}$<br>$k^{1883}$<br>$v^{1861}$ | $i^{1940}$<br>$w^{1816}$<br>$j^{1913}$<br>$k^{1884}$<br>$v^{1862}$ | $p^{2019}$ |
| | $f^{1998}$<br>$j^{1922}$<br>$v^{1871}$ | $f^{2003}$<br>$j^{1921}$<br>$v^{1870}$ | $h^{1961}$<br>$w^{1823}$<br>$i^{1947}$<br>$j^{1920}$<br>$v^{1869}$ | $h^{1966}$<br>$w^{1822}$<br>$i^{1946}$<br>$j^{1919}$<br>$v^{1868}$ | $h^{1973}$<br>$w^{1821}$<br>$i^{1945}$<br>$j^{1918}$<br>$v^{1867}$ | $h^{1978}$<br>$w^{1820}$<br>$i^{1944}$<br>$j^{1917}$<br>$k^{1888}$<br>$v^{1866}$ | $i^{1943}$<br>$w^{1819}$<br>$j^{1916}$<br>$k^{1887}$<br>$v^{1865}$ | $i^{1942}$<br>$w^{1818}$<br>$j^{1915}$<br>$k^{1886}$<br>$v^{1864}$ | $i^{1941}$<br>$w^{1817}$<br>$j^{1914}$<br>$k^{1885}$<br>$v^{1863}$ | $p^{2020}$ |
| | $f^{1999}$<br>$j^{1923}$<br>$k^{1889}$<br>$v^{1872}$ | $f^{2002}$<br>$j^{1924}$<br>$k^{1890}$<br>$v^{1873}$ | $h^{1962}$<br>$w^{1824}$<br>$i^{1948}$<br>$j^{1925}$<br>$k^{1891}$ | $h^{1965}$<br>$w^{1825}$<br>$i^{1949}$<br>$j^{1926}$<br>$k^{1892}$ | $h^{1974}$<br>$w^{1826}$<br>$i^{1950}$<br>$j^{1927}$<br>$k^{1893}$ | $h^{1977}$<br>$w^{1827}$<br>$i^{1951}$<br>$j^{1928}$<br>$k^{1894}$ | $i^{1952}$<br>$w^{1828}$<br>$j^{1929}$<br>$k^{1895}$ | $i^{1953}$<br>$w^{1829}$<br>$j^{1930}$<br>$k^{1896}$ | $i^{1954}$<br>$w^{1830}$<br>$j^{1931}$<br>$k^{1897}$ | $p^{2021}$ |
| | $f^{2000}$<br>$k^{1906}$ | $f^{2001}$<br>$k^{1905}$ | $h^{1963}$<br>$k^{1904}$<br>$w^{1837}$ | $h^{1964}$<br>$k^{1903}$<br>$w^{1836}$ | $h^{1975}$<br>$i^{1959}$<br>$k^{1902}$<br>$w^{1835}$ | $h^{1976}$<br>$i^{1958}$<br>$k^{1901}$<br>$w^{1834}$ | $i^{1957}$<br>$j^{1934}$<br>$k^{1900}$<br>$w^{1833}$ | $i^{1956}$<br>$j^{1933}$<br>$k^{1899}$<br>$w^{1832}$ | $i^{1955}$<br>$j^{1932}$<br>$k^{1898}$<br>$w^{1831}$ | $p^{2022}$ |

**Fig.23. Period 21**

| | | | | | | | | | | |
|---|---|---|---|---|---|---|---|---|---|---|
| $s^{2023}$ | | | | | | | | | | |
| $s^{2024}$ | $d^{2025}$<br>$f^{2026}$ | $d^{2296}$<br>$f^{2295}$ | $d^{2297}$ | $d^{2298}$ | $d^{2299}$ | $d^{2300}$ | $d^{2301}$ | $d^{2302}$ | $d^{2303}$ | $d^{2304}$ |
| | $f^{2027}$<br>$g^{2028}$<br>$v^{2036}$ | $f^{2294}$<br>$t^{2039}$<br>$g^{2029}$<br>$v^{2037}$ | $h^{2030}$<br>$t^{2040}$<br>$g^{2268}$<br>$v^{2126}$<br>$w^{2038}$ | $h^{2257}$<br>$t^{2041}$<br>$g^{2269}$<br>$v^{2127}$<br>$w^{2085}$ | $h^{2258}$<br>$t^{2042}$<br>$g^{2270}$<br>$v^{2128}$<br>$w^{2086}$ | $g^{2271}$<br>$k^{2035}$<br>$t^{2043}$<br>$v^{2129}$<br>$w^{2087}$ | $g^{2272}$<br>$k^{2162}$<br>$t^{2044}$<br>$v^{2130}$<br>$w^{2088}$ | $g^{2273}$<br>$k^{2163}$<br>$t^{2045}$<br>$v^{2131}$<br>$w^{2089}$ | $g^{2274}$<br>$k^{2164}$<br>$t^{2046}$<br>$v^{2132}$<br>$w^{2090}$ | $p^{2305}$ |
| | $f^{2284}$<br>$g^{2283}$<br>$v^{2141}$ | $f^{2293}$<br>$t^{2054}$<br>$g^{2282}$<br>$v^{2140}$ | $h^{2031}$<br>$t^{2053}$<br>$g^{2281}$<br>$v^{2139}$<br>$w^{2097}$ | $h^{2256}$<br>$t^{2052}$<br>$g^{2280}$<br>$v^{2138}$<br>$w^{2096}$ | $h^{2259}$<br>$t^{2051}$<br>$g^{2279}$<br>$v^{2137}$<br>$w^{2095}$ | $g^{2278}$<br>$k^{2168}$<br>$t^{2050}$<br>$v^{2136}$<br>$w^{2094}$ | $g^{2277}$<br>$k^{2167}$<br>$t^{2049}$<br>$v^{2135}$<br>$w^{2093}$ | $g^{2276}$<br>$k^{2166}$<br>$t^{2048}$<br>$v^{2134}$<br>$w^{2092}$ | $g^{2275}$<br>$k^{2165}$<br>$t^{2047}$<br>$v^{2133}$<br>$w^{2091}$ | $p^{2306}$ |
| | $f^{2285}$<br>$j^{2033}$<br>$v^{2142}$ | $f^{2292}$<br>$t^{2055}$<br>$j^{2034}$<br>$v^{2143}$ | $h^{2248}$<br>$w^{2098}$<br>$t^{2056}$<br>$i^{2032}$<br>$j^{2195}$<br>$v^{2144}$ | $h^{2255}$<br>$w^{2099}$<br>$t^{2057}$<br>$i^{2223}$<br>$j^{2196}$<br>$v^{2145}$ | $h^{2260}$<br>$w^{2100}$<br>$t^{2058}$<br>$i^{2224}$<br>$j^{2197}$<br>$v^{2146}$ | $h^{2267}$<br>$w^{2101}$<br>$t^{2059}$<br>$i^{2225}$<br>$j^{2198}$<br>$k^{2169}$<br>$v^{2147}$ | $i^{2226}$<br>$w^{2102}$<br>$t^{2060}$<br>$j^{2199}$<br>$k^{2170}$<br>$v^{2148}$ | $i^{2227}$<br>$w^{2103}$<br>$t^{2061}$<br>$j^{2200}$<br>$k^{2171}$<br>$v^{2149}$ | $i^{2228}$<br>$w^{2104}$<br>$t^{2062}$<br>$j^{2201}$<br>$k^{2172}$<br>$v^{2150}$ | $p^{2307}$ |
| | $f^{2286}$<br>$j^{2210}$<br>$v^{2159}$ | $f^{2291}$<br>$t^{2070}$<br>$j^{2209}$<br>$v^{2158}$ | $h^{2249}$<br>$w^{2111}$<br>$t^{2069}$<br>$i^{2235}$<br>$j^{2208}$<br>$v^{2157}$ | $h^{2254}$<br>$w^{2110}$<br>$t^{2068}$<br>$i^{2234}$<br>$j^{2207}$<br>$v^{2156}$ | $h^{2261}$<br>$w^{2109}$<br>$t^{2067}$<br>$i^{2233}$<br>$j^{2206}$<br>$v^{2155}$ | $h^{2266}$<br>$w^{2108}$<br>$t^{2066}$<br>$i^{2232}$<br>$j^{2205}$<br>$k^{2176}$<br>$v^{2154}$ | $i^{2231}$<br>$w^{2107}$<br>$t^{2065}$<br>$j^{2204}$<br>$k^{2175}$<br>$v^{2153}$ | $i^{2230}$<br>$w^{2106}$<br>$t^{2064}$<br>$j^{2203}$<br>$k^{2174}$<br>$v^{2152}$ | $i^{2229}$<br>$w^{2105}$<br>$t^{2063}$<br>$j^{2202}$<br>$k^{2173}$<br>$v^{2151}$ | $p^{2308}$ |
| | $f^{2287}$<br>$j^{2211}$<br>$k^{2177}$<br>$v^{2160}$ | $f^{2290}$<br>$t^{2071}$<br>$j^{2212}$<br>$k^{2178}$<br>$v^{2161}$ | $h^{2250}$<br>$w^{2112}$<br>$t^{2072}$<br>$i^{2236}$<br>$j^{2213}$<br>$k^{2179}$ | $h^{2253}$<br>$w^{2113}$<br>$t^{2073}$<br>$i^{2237}$<br>$j^{2214}$<br>$k^{2180}$ | $h^{2262}$<br>$w^{2114}$<br>$t^{2074}$<br>$i^{2238}$<br>$j^{2215}$<br>$k^{2181}$ | $h^{2265}$<br>$w^{2115}$<br>$t^{2075}$<br>$i^{2239}$<br>$j^{2216}$<br>$k^{2182}$ | $i^{2240}$<br>$w^{2116}$<br>$t^{2076}$<br>$j^{2217}$<br>$k^{2183}$ | $i^{2241}$<br>$w^{2117}$<br>$t^{2077}$<br>$j^{2218}$<br>$k^{2184}$ | $i^{2242}$<br>$w^{2118}$<br>$t^{2078}$<br>$j^{2219}$<br>$k^{2185}$ | $P^{2309}$ |
| | $f^{2288}$<br>$k^{2194}$ | $f^{2289}$<br>$k^{2193}$ | $h^{2251}$<br>$k^{2192}$<br>$w^{2125}$ | $h^{2252}$<br>$t^{2084}$<br>$k^{2191}$<br>$w^{2124}$ | $h^{2263}$<br>$t^{2083}$<br>$i^{2247}$<br>$k^{2190}$<br>$w^{2123}$ | $h^{2264}$<br>$t^{2082}$<br>$i^{2246}$<br>$k^{2189}$<br>$w^{2122}$ | $i^{2245}$<br>$t^{2081}$<br>$j^{2222}$<br>$k^{2188}$<br>$w^{2121}$ | $i^{2244}$<br>$t^{2080}$<br>$j^{2221}$<br>$k^{2187}$<br>$w^{2120}$ | $i^{2243}$<br>$t^{2079}$<br>$j^{2220}$<br>$k^{2186}$<br>$w^{2119}$ | $p^{2310}$ |

**Fig.24. Period 22**

### Experimental results suggesting existence of superheavy elements

In this paragraph we briefly consider experimental data that may be evidence of superheavy elements existence with relatively long half-decay times sufficient for their discovery and experimental investigation of their properties. Unfortunately only mass-spectra (and similar experimental results determining only atomic mass) of the specimens are available, therefore their chemical properties and electron structure remain unknown.

Fig.25-27 presents experimental data mass spectra obtained, obtained by laboratory of electrodynamics investigations "PROTON-21" group (Ukraine, Kyiv) by initiating an electron collapse wave in the specimen. As a result, the nuclei locally transform into collective state with following generation of majority of stable light nuclei and also some amount of super-heavy elements.

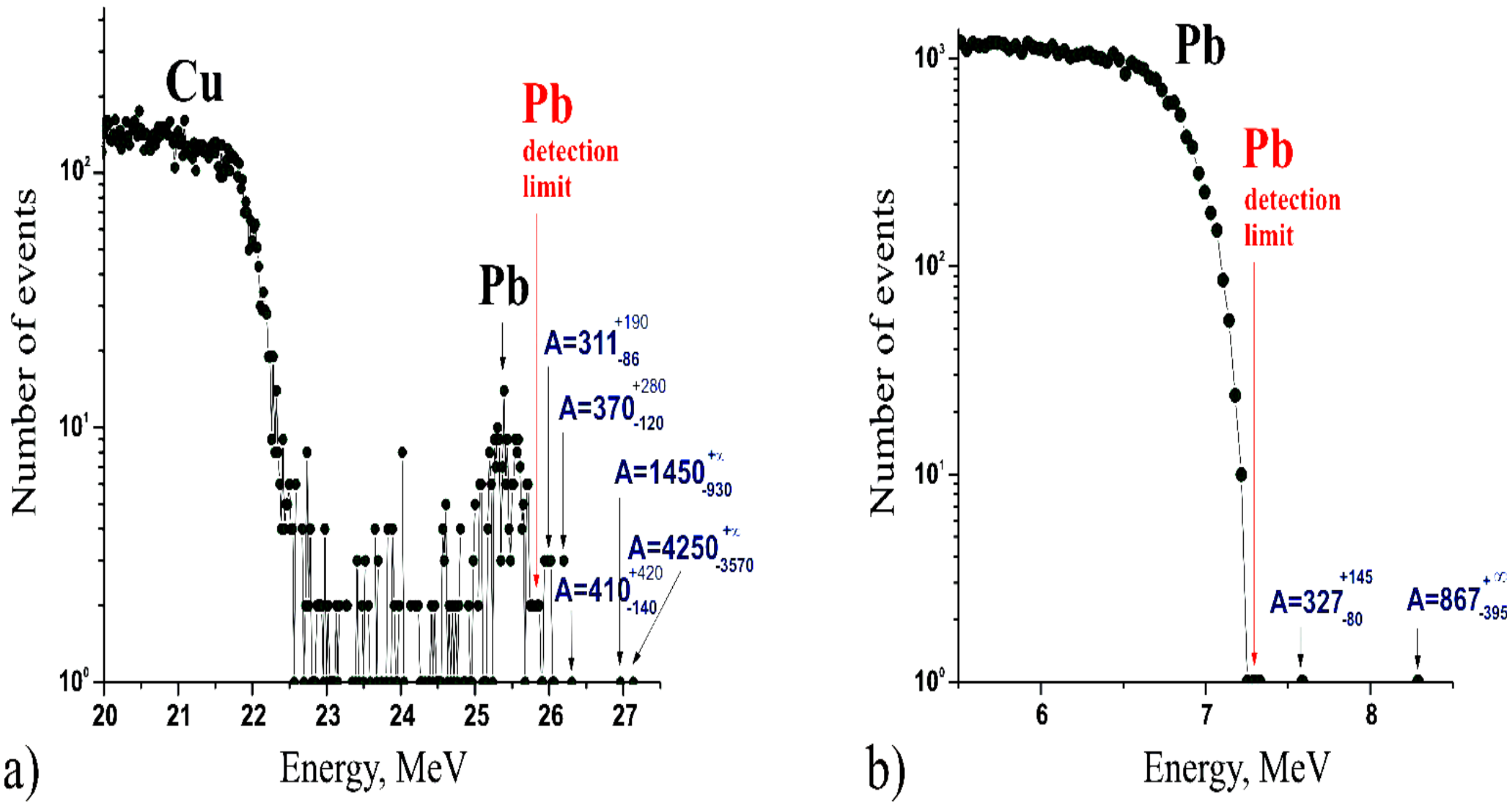

**Fig. 25.** 7 unidentified atomic masses events (atomic masses 311, 327, 370, 410, 867, 1450 and 4250) with accuracy of 1-3 events [37].

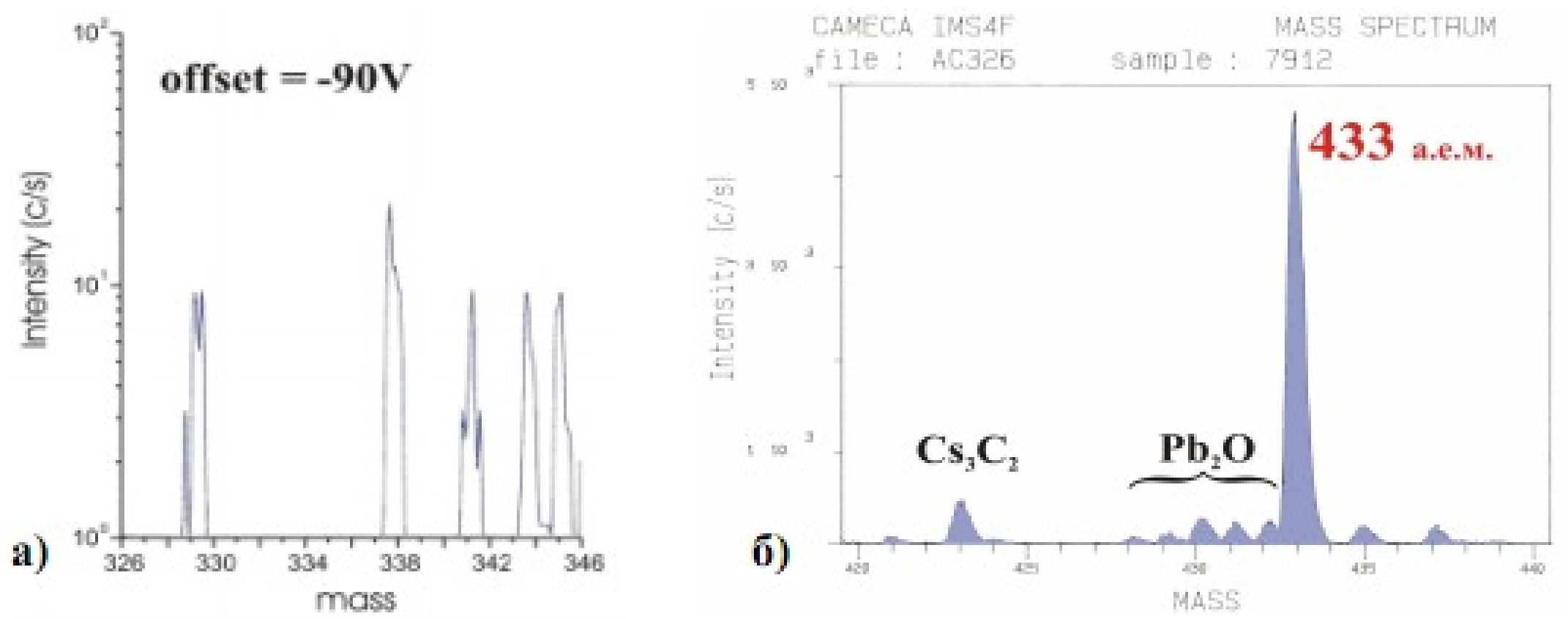

**Fig. 26.** Secondary ion mass-spectrometry (IMS-4f, masses analyzed up to 500). 6 unidentified atomic masses: 329.5, 338, 341.5, 343.5, 344, 345.5 and 433 [37-39].

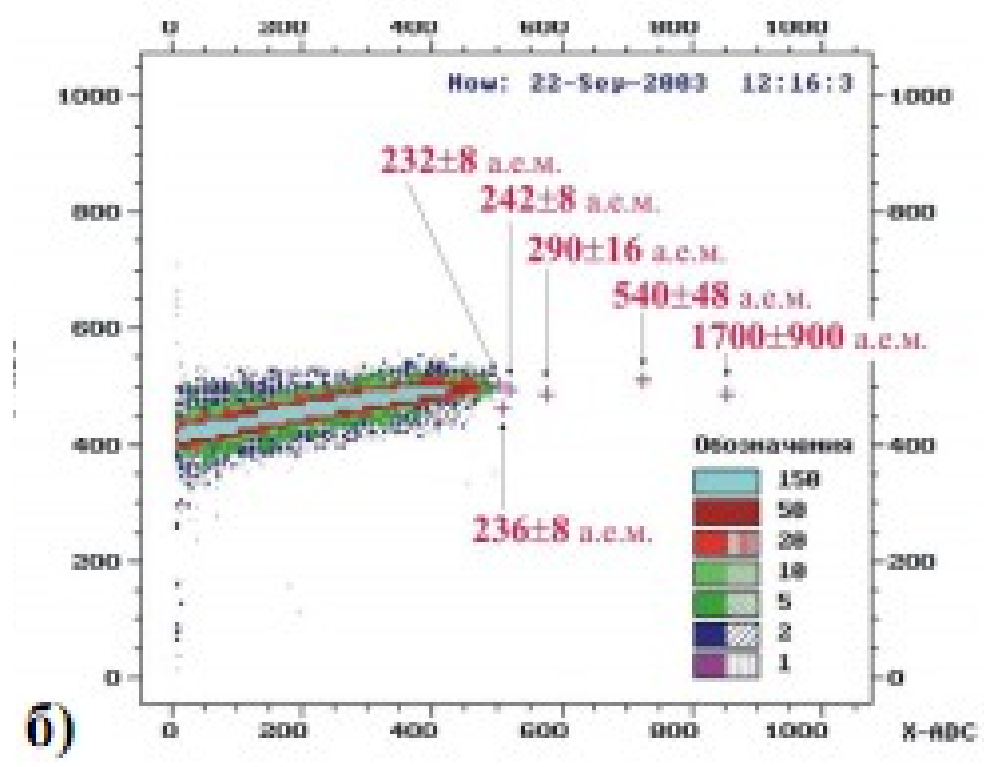

**Fig. 27**. Scattering nitrogen ($^{14}$N) ions by 150$^{o}$ angle with primary energy 8.7 MeV (sample № 8163). There are 6 unidentified atomic masses: 232, 236, 242, 290, 540 and 1700 [39].

Another set of experiments my Marinov [1, 40] describes superheavy elements found in natural metals. In this experiment the metals were «enriched» with superheavy elements by prolonged sublimation vaporization. Therefore according to the Author claims the following stable superheavy elements were discovered: Z=111 in Au and Z=122 in Th.

Let's sum up all these elements in table 2. As we have already discussed above, the atomic mass to atomic number conversion is a very ambiguous procedure. In this table we've used static model with coefficient 2.54. In such case this table is more an example of superheavy atoms identification than some exact calculations. However it shows well the possibilities of the model.

Table 3 shows the same information for confirmed synthesized elements with atomic numbers 105-118.

The element type and analogue is determined according to the spatial periodic table.

**Table 2. Candidates for super heavy elements identification.**

| No | Atomic mass | Charge (atomic number) | Type | Analogue | Notes |
|---|---|---|---|---|---|
| 1 | 292 | 122 | f | Th | Found by Marinov in natural Th sample |
| | 311 | | | | Proton-21 |
| 2 | 327 | 129 | g | Fe | Proton-21 |
| 3 | 329.5 | 130 | g | Co | Proton-21 |
| | 329.9 | | | | Proton-21 |
| 4 | 338 | 133 | g | Cu | Proton-21 |
| 5 | 341.5 | 134 | g | Ni | Proton-21 |
| 6 | 343.5 | 135 | g | Co | Proton-21 |
| | 344 | | | | Proton-21 |
| 7 | 345.2 | 136 | g | Fe | Proton-21 |
| 8 | 370 | 146 | f | Gd | Proton-21 |
| 9 | 410 | 161 | d | Cu | Proton-21 |
| 10 | 433 | 170 | s | Ba | Proton-21 |
| 11 | 540 | 213 | p | Tl | Proton-21 |
| 12 | 867 | 341 | f | Tb | Proton-21 |
| 13 | 1450 | 571 | j | V | Proton-21 |
| 14 | 1700 | 669 | f | Tm | Proton-21 |
| 15 | 4250 | 1673 | j | Ni | Proton-21 |

**Table 3. Identification of known superheavy elements.**

| Charge (atomic number) | Type | Analogue | Notes |
|---|---|---|---|
| 105 | d | Ta | |
| 106 | d | W | |
| 107 | d | Re | |
| 108 | d | Os | >100 atoms synthesized |
| 109 | d | Ir | |
| 110 | d | Pt | |
| 111 | d | Au | Found by Marinov in Au sample |
| 112 | d | Hg | ~ 75 atoms synthesized |
| 113 | p | Tl | |
| 114 | p | Pb | ~ 80 atoms synthesized; Synthesized by Proton-21 |
| 115 | p | Bi | ~ 50 atoms synthesized |
| 116 | p | Po | ~ 35 atoms synthesized |
| 117 | p | At | 2 successful synthesis events (yet unconfirmed) |
| 118 | p | Rn | 3 successful synthesis events |

## Conclusions

1. This paper presents new spatial periodic table that includes 2310 elements and has 22 periods.
2. Possible ways of search for and identification of stable superheavy elements are described.
3. Analysis results of 15 superheavy elements with atomic numbers 122, 129, 130, 133, 134, 135, 136, 146, 161, 170, 213, 341, 571, 669 and 1673 are given. Based on spatial periodic table the recently-synthesized superheavy elements and superheavy elements reported by A.Marinov and Proton-21 laboratory chemical properties were determined.